\documentclass[10pt,twocolumn,preprintnumbers,amsmath,amssymb,nofootinbib
,superscriptaddress,aps,prd]{revtex4} 
\usepackage{graphicx,amsmath,amssymb,amstext}
\usepackage{amsbsy,amsfonts,amsthm,color}
\usepackage{xcolor}
\usepackage[colorlinks,linkcolor=red,citecolor=blue,urlcolor=blue ]{hyperref}
\newcommand{\covar}{\mathsf{\Sigma}}
\newcommand{\mnras}{{\em Mon.\ Not.\ R.\ Astron.\ Soc.}}
\newcommand{\jcap}{JCAP}
\newcommand{\ssr}{Space Science Reviews}
\newcommand{\physrep}{Physics Reports}
\newcommand{\aap}{Astronomy and Astrophysics}

\begin{document}

\title{The effect on cosmological parameter estimation of a parameter-dependent covariance matrix}

\author{Darsh Kodwani}
\email{darsh.kodwani@physics.ox.ac.uk}
\affiliation{Department of Physics, University of Oxford, Denys Wilkinson Building, Keble Road, Oxford, OX1 3RH, UK}

\author{David Alonso}
\email{david.alonso@physics.ox.ac.uk}
\affiliation{School of Physics and Astronomy, Cardiff University, The Parade, Cardiff, CF24 3AA, UK}
\affiliation{Department of Physics, University of Oxford, Denys Wilkinson Building, Keble Road, Oxford, OX1 3RH, UK}

\author{Pedro G. Ferreira}
\email{pedro.ferreira@physics.ox.ac.uk}
\affiliation{Department of Physics, University of Oxford, Denys Wilkinson Building, Keble Road, Oxford, OX1 3RH, UK}

  \begin{abstract}
    Cosmological large-scale structure analyses based on two-point correlation functions often assume a Gaussian likelihood function with a fixed covariance matrix. We study the impact on cosmological parameter estimation of ignoring the parameter dependence of this covariance matrix, focusing on the particular case of joint weak-lensing and galaxy clustering analyses. Using a Fisher matrix formalism (calibrated against exact likelihood evaluation in particular simple cases), we quantify the effect of using a parameter dependent covariance matrix on both the bias and variance of the parameters. We confirm that the approximation of a parameter-independent covariance matrix is exceptionally good in all realistic scenarios. The information content in the covariance matrix (in comparison with the two point functions themselves) does not change with the fractional sky coverage. Therefore the increase in information due to the parameter dependent covariance matrix becomes negligible as the number of modes increases. Even for surveys covering less than 1\% of the sky, this effect only causes a bias of up to ${\cal O}(10\%)$ of the statistical uncertainties, with a misestimation of the parameter uncertainties at the same level or lower. The effect will only be smaller with future large-area surveys. Thus for most analyses the effect of a parameter-dependent covariance matrix can be ignored both in terms of the accuracy and precision of the recovered cosmological constraints. 
  \end{abstract}

\maketitle
  \section{Introduction}\label{sec:intro}
    We are entering a phase in which cosmological galaxy surveys will have remarkable constraining power. This arises from the fact that they will cover large areas of the sky, to large depths and, in consequence, with a high number density \cite{2009arXiv0912.0201L, 2011arXiv1110.3193L,2013arXiv1305.5422S,2017arXiv170801530D,2018MNRAS.474.4894J,2018arXiv180909148H}. As a result, the statistical power of these surveys will be astounding. To access their scientific potential, it will be necessary to control any systematic effects with exquisite accuracy. While there has been much focus on the instrumental and astrophysical limitations of any particular survey, care is now being given to the analysis pipeline and the approximations which are currently being used. In this paper we focus on one such aspect: the covariance matrix in the likelihood analysis used for constraining cosmological parameters.

    Given a set of cosmological observations, for example in the form of a map, a galaxy catalogue, a correlation function or power spectrum, the process for estimating cosmological parameters is straightforward. One feeds the data into a likelihood function which assesses the likelihood of a given theory, given the data. With a judicious use of Bayes theorem, one explores a range of theoretical parameters (that can be cosmological but also can characterise underdetermined properties of the survey or astrophysical effects that may be present) to find the subspace of parameters which is best compatible with the data. In doing so, one also derives uncertainties on these parameters. It is therefore equally important for this procedure to produce unbiased parameter estimates and precise error bars.
    
    There are a number of effects that can plague cosmological parameter estimation. For a start there are errors due to the inaccurate modelling of simulated data, $x(p)$ and the corresponding, covariance matrix $\covar$ estimated from them. These can come from not understanding the underlying model (e.g. baryonic effects in the matter power spectrum \cite{PhysRevD.87.043509, 0004-637X-672-1-19, vdb,2015JCAP...12..049S,2018MNRAS.480.3962C}, the precise nature of the galaxy-matter connection \citep{0004-637X-734-2-94,PhysRevD.81.083509} or the impact of intrinsic alignments \cite{2015SSRv..193....1J,2015PhR...558....1T}). Then there are errors due to incorrect assumptions about the statistics of the data and covariance matrix. For example there may be non-Gaussian corrections to $\covar$ (i.e. corrections associated to the non-Gaussian nature of the fields being correlated). Then there are errors due to an incorrect model of the likelihood (e.g. assuming a Gaussian function of the data vector by default). Finally, even within the Gaussian approximation, there are errors that may arise from ignoring the parameter dependence of the covariance matrix. It is on this last source of errors that we will focus in this paper.

    There is a growing literature on the accuracy of likelihood functions and its impact on parameter estimation. A key focus has been on how well one needs to know the covariance matrix and how Gaussian the likelihood function is. Covariance matrices are generally estimated from a large number of simulations; there is now a clear understanding of the errors associated with such a process and how to correct for them \cite{Morrison:2013tqa, Hall:2018umb, Dodelson:2013uaa, OConnell:2018oqr, Taylor:2014ota, Joachimi:2014hca, Paz:2015kwa, Pearson:2015gca, Petri:2016wlu} . A recent proposal in \cite{Heavens:2017efz} provides a novel data compression algorithm that reduces the number of simulations needed by a factor of 1000 to estimate the covariance matrix for Gaussianly distributed data. There have also been attempts at constructing better models, or approximations, for the likelihood function that include the fact that it may be non-Gaussian \cite{Sellentin:2015waz,Sellentin:2016psv}. 

    There have been attempts to analytically calculate the non-Gaussian contributions to the covariance matrix. The authors of \cite{Mohammed:2016sre} use a perturbation theory approach to compute the non-Gaussian contribution of the matter power spectrum to 1-loop and compare the results to simulations. A linear response approach to calculating the non-Gaussian contribution of the covariance matrix is presented in \cite{Barreira:2017sqa,Barreira:2017kxd}. In the case of large-scale structure observables like the cosmic shear power spectrum, it has been pointed out that that the dominant contribution to the non-Gaussian covariance is the term that comes from super-sample covariance (SSC) \cite{Barreira:2018jgd}. Separate-universe simulations have been used to evaluate the response functions for various observables such as weak lensing and the matter power spectrum \cite{Li:2015jsz,Wagner:2015gva}.

    In this paper, our aim is to quantify the impact on parameter estimation of including (or not) the parameter dependence of the covariance matrix for upcoming photometric redshift surveys. Previous work on this topic has focused on the Fisher information of a parameter-dependent covariance in the two-point likelihood of Gaussian random fields \cite{2013A&A...551A..88C}, and on the overall parameter dependence of the cosmic shear two-point covariance \cite{Eifler:2008gx}, where the authors quantified the effect on the likelihood contours of $\sigma_8$ and $\Omega_m$, using both analytic and ray tracing simulations. Here we will focus on galaxy clustering and galaxy shear for a tomographic survey, and we will systematically analyse the effect of a parameter-dependent covariance matrix (including the effects of super-sample covariance) in terms of information content, parameter biases and final uncertainties.

    The paper is structured as follows. In Section \ref{sec:theory} we present the various types of likelihoods which we will be studying and show how we can use the Fisher forecast formalism to find an estimate of the bias and errors of the model parameters; we consider a few simplified models to get an idea of what one should expect in the more general, realistic case. In Section \ref{sec:forecasts} we look at a realistic survey scenario, involving a combination of weak lensing and galaxy clustering and calculate the bias and compare the errors depending on the choice of likelihood.  In Section \ref{sec:summary} we discuss the implications of our results while in Appendix A we present a few, key, technical aspects of our calculation of the covariance matrix. 

\section{Approximating the likelihood}\label{sec:theory}

The aim of this section we quantify the impact of a parameter-dependent covariance on the bias and variance of parameter estimates. The impact of biases on the data vector has been quantified in e.g. \cite{2008MNRAS.391..228A,2009MNRAS.392.1153T}, and the parameter dependence of the covariance has been explored in \cite{Eifler:2008gx,2017MNRAS.465.4016R}. In particular we calculate the bias in inferred parameters explicitly due to the parameter dependence of the covariance matrix, which is a key result that hasn't been calculated before as far as we know.

In general, for a Gaussian data vector ${\bf d}$, the likelihood is given by
    \begin{equation}\label{eq:multinorm}
      p({\bf d}|\vec{\theta})=\frac{\exp\left[-\frac{1}{2}\left({\bf d}-{\bf t}\right)^T\,\covar^{-1}\left({\bf d}-{\bf t}\right)\right]}{\sqrt{{\rm det}\left(2\pi\,\covar\right)}},
    \end{equation}
    where both the mean ${\bf t}$ and the covariance matrix $\covar$ may in principle depend on the parameters ${\vec \theta}$.
    
    To begin with, we will start with the simple scenario, well known in the literature, in which the data is a single Gaussian sky map. It will be instructive to review some basic points about Gaussian likelihoods and the role covariance matrices play in them. In doing so, we will identify two approximations to the true likelihood function (for parameter-dependent and independent covariances) and how one might assess the difference in the parameter estimates they lead to.
 
    \subsection{Likelihoods and covariances: the case of a 2-dimensional Gaussian field}\label{ssec:theory.single_map}
      Consider the harmonic coefficients $a_{\ell m}$ of a single full-sky map. Under the assumption that they are statistically isotropic, their covariance matrix is diagonal, and given by the angular power spectrum $C_\ell$:
      \begin{equation}
        \left\langle a_{\ell m}\,a^*_{\ell' m'}\right\rangle\equiv C_\ell\,\delta_{\ell\ell'}\delta_{m m'}.
      \end{equation}
      If we further assume that the mean of the map is zero, and that it is Gaussianly distributed, its likelihood is completely determined by the power spectrum, and takes the form:
      \begin{equation}
        p(a_{\ell m}|{\vec \theta})=\prod_\ell\exp\left[-\frac{2\ell+1}{2}\frac{C^e_\ell}{C_\ell}\right]\left(2\pi C_\ell\right)^{-\frac{2\ell+1}{2}},
      \end{equation}
      where $C^e_\ell\equiv\sum_{m=-\ell}^\ell|a_{\ell m}|^2/(2\ell+1)$ is the estimated power spectrum. For simplicity, let us focus on a single multipole order $\ell$, and we will also use the goodness of fit $G=-2\log p$. Assuming flat priors on all parameters, the posterior distribution for $\vec{\theta}$ is simply given by this likelihood, and therefore
      \begin{equation}
        G(\vec{\theta}|a_{\ell m})=(2\ell+1)\left[\frac{C^e_\ell}{C_\ell}+\log(C_\ell)\right]+{\rm const.}
      \end{equation}

      It is often desirable to compress the data into a quadratic summary statistic, such as $C^e_\ell$. If we consider $C^e_\ell$ to be the data, then the likelihood is described by the Wishart distribution \cite{Hamimeche:2008ai} given by
      \begin{align}
        G_{\rm exact}({\vec \theta}|C^e_\ell)=&(2\ell+1)\left[\frac{C^e_\ell}{C_\ell}+\log C_\ell-\frac{2\ell-1}{2\ell+1}\ln(C^e_\ell)\right] \nonumber \\& +{\rm constant}, \label{eq:fulllike2D}
      \end{align}
      where we have used the label ``exact'' to distinguish this distribution from the two approximations below. The extra term in this expression is a Jacobian/volume term that comes about due to the change of variables from $a^e_{\ell m}$ to $C^e_{\ell}$. This expression effectively tells us that $z=(2\ell+1){C}^e_\ell/C_\ell$ obeys a $\chi^2_{2\ell+1}$ distribution.
      
      For any distribution $G$, the Fisher information matrix is given by
      \begin{eqnarray}
        {\cal F}_{\mu\nu}=\frac{1}{2}\left\langle \partial_\mu\partial_\nu G\right\rangle,
      \end{eqnarray}
      where we have used the shorthand $\partial_\mu=\partial/\partial \theta^\mu \equiv \partial / \partial \vec{\theta}$. For the exact likelihood in Eq. \ref{eq:fulllike2D} we find
      \begin{eqnarray}
        {\cal F}^{\rm exact}_{\mu\nu}=\frac{2\ell+1}{2}\frac{\partial_\mu C_\ell}{C_\ell}\frac{\partial_\nu C_\ell}{C_\ell}
      \end{eqnarray}

      Until now, all our calculations have been exact in the limit of Gaussian, full-sky maps. Let us now consider a simplification that can be made if we assume that $C^e_\ell$ itself is Gaussianly distributed. This is a valid approximation in the large-$\ell$ limit, since the number of independent modes contributing to a given $C_\ell$ grows like $\ell$, and the Central Limit Theorem (CLT) eventually applies. In this approximation, and under the assumption that the covariance is parameter-independent, the likelihood takes the form
      \begin{eqnarray}
        G_{\rm PI}({\vec p}|C^e_\ell)\simeq {\tilde G}({\vec p}|C^e_\ell)=\frac{(C^e_\ell-C_\ell)^2}{\covar^f_\ell},
      \end{eqnarray}
      where the label PI identifies this distribution with the case of a ``parameter-independent'' covariance matrix, and where we have defined the power spectrum covariance $\covar^f_\ell$
      \begin{equation}
        \langle C^e_\ell C^e_{\ell'}\rangle-\langle C^e_\ell\rangle\langle C^e_{\ell'}\rangle\equiv\covar^f_\ell \delta_{\ell\ell'}.
      \end{equation}
      Wick's theorem can be used to relate $\covar^f_\ell$ to the underlying fiducial power spectrum $C^f_\ell$, finding the well-known result
      \begin{equation}
        \covar^f_\ell=\frac{2C^{f\,2}_\ell}{2\ell+1}
      \end{equation}
      where $C^f_\ell$ is  a fiducial power spectrum which is {\it independent} of the cosmological parameters. One can show that this approximation can be obtained by Taylor expanding Eq \ref{eq:fulllike2D} in $\xi=(C_\ell-C^e_\ell)/C^e_\ell$ to get 
      \begin{align}
        {\tilde G}({\vec p}|C^e_\ell)=&(2\ell+1)\left[\frac{1}{2}\xi^2+\frac{2}{2\ell+1}\log C^e_\ell\right] \nonumber \\ &+{\rm constant}
      \end{align}
      and assuming $C^e_\ell\simeq C^f_\ell$.
      
      This approximate likelihood is similar to the original exact likelihood in two ways. First, their Fisher matrices coincide if $C^f_\ell$ is the true underlying power spectrum
      \begin{equation}
        {\cal F}^{\rm PI}_{\mu\nu}=\frac{2\ell+1}{2}\frac{\partial_\mu C_\ell}{C^f_\ell}\frac{\partial_\nu C_\ell}{C^f_\ell}.
      \end{equation}
      Secondly, in the simplest case, where our only free parameter is the amplitude of the power spectrum itself (i.e. $\theta\equiv C_\ell$), both likelihoods yield unbiased maximum likelihood estimators for this parameter:
      \begin{equation}
        \hat{C}^{\rm exact}_\ell=\hat{C}^{\rm PI}_\ell=C^e_\ell,\hspace{12pt}\langle C^e_\ell\rangle=C^f_\ell
      \end{equation}

      Consider now a different approach and take the large $\ell$ limit of Eq (\ref{eq:fulllike2D}). Applying the CLT we have that $\xi$ obeys a Normal distribution, ${\cal N}[0,\ell+1/2]$, i.e.:
      \begin{eqnarray}
        G_{\rm PD}({\vec \theta}|C_\ell)=\log(\covar_\ell)+\frac{(C_\ell-C^e_\ell)^2}{\covar_\ell} \label{gaussparam}
      \end{eqnarray}
      where now the covariance matrix
      \begin{eqnarray}
        \covar_\ell=\frac{2}{2\ell+1}C^{2}_\ell \label{2DGcov}
      \end{eqnarray}
      depends on $\theta$ (and hence the label ``PD''). Accounting for this parameter dependence, the Fisher matrix for this distribution is
      \begin{equation}
        {\cal F}^{\rm PD}_{\mu\nu}=\frac{2\ell+5}{2}\frac{\partial_\mu C_\ell}{C_\ell}\frac{\partial_\nu C_\ell}{C_\ell},
      \end{equation}
      and the maximum likelihood estimator for the power spectra is
      \begin{align}
        \hat{C}^{\rm PD}_\ell&=C^e_\ell\frac{2\ell+1}{4}\left[\sqrt{1+\frac{8}{2\ell+1}}-1\right]\\\label{eq:bias_single}
                             &\simeq C^e_\ell\left[1-\frac{2}{2\ell+1}\right],
      \end{align}
      where in the second line we have kept only the first-order term of the large-$\ell$ expansion of the first line. Therefore, $G^{\rm PD}$ reduces to $G^{\rm exact}$ in the large-$\ell$ limit.
      
      We thus see that in this particular case, while assuming a parameter {\it independent} covariance matrix may lead to unbiased parameter estimates, it necessarily leads to a mis-estimate of the parameter uncertainties (unless the chosen fiducial covariance $\covar^f_\ell$ is the underlying true one). A parameter {\it dependent} covariance matrix (assuming a Gaussian approximation for the likelihood) leads to both biased parameter estimates and a misestimate of the uncertainties but there is a well defined limit in which it recovers both correctly and this limit is set by the number of modes being considered in the analysis. This point was made by \cite{2013A&A...551A..88C}, who also show that using a parameter-dependent covariance when approximating the two-point likelihood of Gaussian random fields is formally incorrect.

      The question then arises: how important, in practice, is the parameter dependence in the process of parameter estimation from current and future data sets? In this paper, we go beyond the simple analysis of 2D full-sky Gaussian fields presented here to consider the case of tomographic analyses with non-Gaussian contributions to the covariance matrix.

    \subsection{Likelihoods and covariances: the general case}\label{ssec:theory.general}
      The aim of this section is to derive approximate expressions for the parameter uncertainties and biases associated to the parameter dependence of the covariance matrix. Using the identity $\log{\rm det}{\sf M}={\rm Tr}\log{\sf M}$, let us start by writing the goodness of fit for the generic multi-variate Gaussian distribution (Eq. \ref{eq:multinorm}) as
      \begin{eqnarray}
        G_{\rm gen}({\bf d}|\vec{\theta})= \left({\bf d}-{\bf t}\right)^T \covar^{-1} \left({\bf d}-{\bf t}\right)+{\rm Tr}\left(\log\covar\right) \label{eq:ggen}.
      \end{eqnarray}
      Let us now define three sets of parameters
      \begin{enumerate}
        \item $\vec{\theta}_{\rm T}$: the true parameters that generate the data.
        \item $\vec{\theta}_{\rm PD}$: the maximum-likelihood parameters found by minimizing $G_{\rm PD}$, the version of $G_{\rm gen}$ in which the covariance matrix \emph{depends} on $\vec{\theta}$.
        \item $\vec{\theta}_{\rm PI}$: the maximum-likelihood parameters found by minimizing $G_{\rm PI}$, the version of $G_{\rm gen}$ in which the covariance matrix \emph{does not depend} on $\vec{\theta}$.
      \end{enumerate}
      $\vec{\theta}_{\rm T}$ generate the data in the sense that $\langle {\bf d}\rangle={\bf t}(\vec{\theta}_{\rm T})$\footnote{We note that this assumes that the theory we have, in this case $\Lambda CDM$, is the \emph{true} theory of the universe. This of course may not be true however answering that question is tangential to the goal of this paper and thus we do not address this further.}  and
      \begin{equation}
        \left\langle [{\bf d}-{\bf t}(\vec{\theta}_{\rm T})]\,[{\bf d}-{\bf t}(\vec{\theta}_{\rm T})]^T\right\rangle=\covar(\vec{\theta}_{\rm T}).
      \end{equation}
      We will also assume that the PI likelihood uses the true covariance as the fiducial one, i.e. $\covar^f=\covar(\vec{\theta}_{\rm T})$\footnote{Note that this choice only simplifies the calculations, but does not affect our results. Choosing any other fiducial point ($\vec{\theta}_{\rm PI}$, or $\vec{\theta}_{\rm PD}$ for example) only leads to second-order corrections.}. This will allow us to isolate the impact of the parameter dependence from the systematics effects associated with using an inaccurate covariance matrix. For brevity, we will often use the shorthand $\covar_{\rm T}\equiv\covar(\vec{\theta}_{\rm T})$. Note that, at this stage, we have not made any statements about the validity of either $G_{\rm PI}$ or $G_{\rm PD}$\footnote{Note however, that as pointed out in \cite{2013A&A...551A..88C} and in Section \ref{ssec:theory.single_map} there is a clear distinction between both for Gaussian random fields, and using a parameter-dependent covariance produces fictitious information.}, and we will focus only on the comparison of their associated parameter uncertainties and on their relative bias.
    
      By definition
      \begin{equation}
        \left.\frac{\partial G_{\rm PD}}{\partial\vec{\theta}}\right|_{\vec{\theta}_{\rm PD}}=0.
      \end{equation}
      Writing $\vec{\theta}_{\rm PD}=\vec{\theta}_{\rm PI}+\Delta\vec{\theta}$ and $G_{\rm PD}=G_{\rm PI}+\Delta G$ we can expand the equation above to linear order, to find
      \begin{equation}
        \left.\frac{\partial^2 G_{\rm PI}}{\partial\vec{\theta}\partial\vec{\theta}}\right|_{\vec{\theta}_{\rm PI}}\Delta\vec{\theta}+\left.\frac{\partial\Delta G}{\partial\vec{\theta}}\right|_{\vec{\theta}_{\rm PI}}=0,
      \end{equation}
      where $\partial/\partial\vec{\theta}$ and $\partial^2/\partial{\vec{\theta}}\partial{\vec{\theta}}$ is shorthand for the parameter gradient and Hessian matrix respctively. Taking the expectation value of the equation above, the parameter bias can be estimated in this approximation as
      \begin{equation}\label{eq:bias_general}
        \Delta\vec{\theta}=-\frac{1}{2}\,\mathcal{F}^{-1}_{\rm PI}\left\langle\left.\frac{\partial \Delta G}{\partial\vec{\theta}}\right|_{\vec{\theta}_{\rm PI}}\right\rangle,
      \end{equation}
      where $\mathcal{F}_{\rm PI}$ is the Fisher matrix for the parameter-independent case, given simply by
      \begin{equation}\label{eq:fisher_PI}
        \mathcal{F}_{{\rm PI},\mu\nu}\equiv\partial_\mu{\bf t}^T\,\covar^{-1}\,\partial_\nu{\bf t}.
      \end{equation}
      To evaluate $\Delta\vec{\theta}$ for the distribution in Eq. \ref{eq:ggen}, let us start by writing $\Delta G$ to first order in $\Delta\covar\equiv\covar-\covar_{\rm T}$
      \begin{align}\nonumber
        \Delta G=&-\left({\bf d}-{\bf t}\right)^T\covar_{\rm T}^{-1}\,\Delta\covar\,\covar_{\rm T}^{-1}\left({\bf d}-{\bf t}\right)+\\&{\rm Tr}\left(\covar_{\rm T}^{-1}\Delta\covar\right).
      \end{align}
      Differentiating with respect to $\vec{\theta}$ we obtain
      \begin{align}\nonumber
        \partial_\mu\Delta G=&2\partial_\mu{\bf t}^T\covar_{\rm T}^{-1}\,\Delta\covar\,\covar_{\rm T}^{-1}({\bf d}-{\bf t})\\\label{eq:dgpre}
                             &+{\rm Tr}\left[\covar^{-1}_{\rm T}\partial_\mu\covar\left(\covar_{\rm T}^{-1}({\bf d}-{\bf t})({\bf d}-{\bf t})^T-1\right)\right],
      \end{align}
      where in the second line we have used the fact that $\partial_\mu\Delta\covar\equiv\partial_\mu\covar$. Before continuing, it is important to note that, according to Eq. \ref{eq:bias_general}, we must evaluate this expression at the parameter-independent best fit $\vec{\theta}_{\rm PI}$. This best fit satisfies
      \begin{equation}
        \left.\frac{\partial G_{\rm PI}}{\partial\theta_\mu}\right|_{\vec{\theta}_{\rm PI}}=-2\partial_\mu{\bf t}^T\covar_{\rm T}^{-1}({\bf d}-{\bf t})=0.
      \end{equation}
      Now let us define $\delta{\bf d}\equiv{\bf d}-{\bf t}(\vec{\theta}_{\rm T})$ and $\delta\vec{\theta}\equiv\vec{\theta}_{\rm PI}-\vec{\theta}_{\rm T}$. To linear order in $\delta\vec{\theta}$, the equation above reads
      \begin{equation}
        \partial_\mu{\bf t}^T\covar_{\rm T}^{-1}(\delta{\bf d}-\partial_\nu{\bf t}\,\delta\theta_\nu)=0,
      \end{equation}
      and we can solve for $\delta\vec{\theta}$ as
      \begin{equation}
        \delta\vec{\theta}=\mathcal{F}_{\rm PI}^{-1}\frac{\partial{\bf t}^T}{\partial\vec{\theta}}\covar^{-1}_{\rm T}\delta{\bf d}
      \end{equation}
      or, equivalently
      \begin{equation}
        {\bf d}-{\bf t}(\vec{\theta}_{\rm PI})=\left[1-\frac{\partial{\bf t}}{\partial\vec{\theta}}\mathcal{F}_{\rm PI}^{-1}\frac{\partial{\bf t}^T}{\partial\vec{\theta}}\covar^{-1}_{\rm T}\right]\delta{\bf d}.
      \end{equation}
      Substituting this in Eq. \ref{eq:dgpre}, making use of the fact that $\langle\delta{\bf d}\rangle=0$ and $\langle\delta{\bf d}\,\delta{\bf d}^T\rangle\equiv\covar_{\rm T}$, and after a little bit of algebra, we obtain
      \begin{equation}
        \left\langle\left.\frac{\partial\Delta G}{\partial\theta_\mu}\right|_{\vec{\theta}_{\rm PI}}\right\rangle=-{\rm Tr}\left[\covar^{-1}_{\rm T}\frac{\partial\covar}{\partial\theta_\mu}\covar_{\rm T}^{-1}\frac{\partial{\bf t}}{\partial\vec{\theta}}\mathcal{F}_{\rm PI}^{-1}\frac{\partial{\bf t}^T}{\partial\vec{\theta}}\right].
      \end{equation}
      With index notation then, the parameter bias is
      \begin{equation}\label{eq:fisher_bias}
        \Delta\theta_\mu=-\frac{1}{2}\mathcal{F}^{-1}_{{\rm PI},\mu\nu}\,\mathcal{F}^{-1}_{{\rm PI},\rho\tau}\,\partial_\rho{\bf t}^T\,\covar^{-1}\,\partial_\nu\covar\,\covar^{-1}\,\partial_\tau{\bf t}.
      \end{equation}
      This is a key result of our paper.
      
      The effect of the parameter-dependent covariance on the final parameter uncertainties can be taken into account simply by accounting for this parameter dependence when deriving the Fisher matrix. Such a calculation  yields \citep{1997ApJ...480...22T}
      \begin{equation}\label{eq:fisher_PD}
        \mathcal{F}_{{\rm PD},\mu\nu}=\mathcal{F}_{{\rm PI},\mu\nu}+\frac{1}{2}{\rm Tr}\left[\partial_\mu\covar\,\covar^{-1}\,\partial_\nu\covar\,\covar^{-1}\right].
      \end{equation}
      The parameter uncertainties can then be estimated by inverting the Fisher matrix in either case.

    \subsection{Large-scale structure likelihoods}\label{ssec:theory.lss}
      Let us now specialise the discussion in the previous section to the case of a data vector made up of the collection of auto- and cross-power spectra between different sky maps, each labelled by a roman index (e.g. $a^i_{\ell m}$ labels the harmonic coefficients of the $i$-th). In general, each sky map will correspond to an arbitrary projected astrophysical field, and the discussion below covers this general case. However, here we will only consider maps from two types of tracers, the galaxy number overdensity $\delta_g$ and the cosmic shear $\gamma$, each measured in a given tomographic redshift bin. The cross-power spectrum between two fields is
      \begin{equation}
        \left\langle a^i_{\ell m}a^{j*}_{\ell'm'}\right\rangle=C^{ij}_\ell\delta_{\ell\ell'}\delta_{mm'}.
      \end{equation}
      $C^{ij}_\ell$ can be related in general to the matter power spectrum in the Limber approximation \cite{2007A&A...473..711S} through \cite{2001PhR...340..291B}
      \begin{equation}\label{eq:pspec_signal}
        C^{ij}_\ell=\int_0^\infty d\chi\,\frac{q^i(\chi)q^j(\chi)}{\chi^2}P\left(z(\chi),k=\frac{\ell+1/2}{\chi}\right),
      \end{equation}
      where the window functions for $\delta_g$ and $\gamma$ are
      \begin{align}
        q^{\delta,i}(\chi)&=b(\chi)\frac{dn^i}{dz}(\chi)\,H(\chi), \label{eq:q_delta}\\
        q^{\gamma,i}(\chi)&=\frac{3H_0^2\Omega_M}{2\,a(\chi)}\int_\chi^{\chi_H} d\chi'\,\frac{dn^i}{dz}(\chi')\,\frac{\chi'-\chi}{\chi\chi'}. \label{eq:q_lensing}
      \end{align}
      Here $b(\chi)$ is the linear galaxy bias, $dn^i/dz$ is the redshift distribution of lens or source galaxies respectively in the $i$-th redshift bin, normalised to 1 when integrated over the full redshift range, and $\chi_H$ is the distance to the horizon.
    
      Since our data vector is an ordered list of cross-power spectra between different pairs of maps $(ij)$ at different scales $\ell$, the covariance matrix depends on 6 indices
      \begin{equation}
        \covar^{ij,\ell}_{mn,\ell'}\equiv\left\langle C^{ij}_\ell\,C^{mn}_{\ell'}\right\rangle.
      \end{equation}	
      In order to account for the non-Gaussian nature of the late-times large-scale structure, we estimate the covariance matrix as a sum of both the Gaussian and super-sample covariance (SSC) contributions \cite{Takada:2013bfn,Mohammed:2016sre}
      \begin{equation}\label{eq:covar_lss}
        \covar=\covar_{\rm G}+\covar_{\rm SSC}.
      \end{equation}
      We neglect other non-Gaussian corrections from the connected trispectrum, which are known to be subdominant with respect to the SSC contribution, at least for cosmic shear studies \cite{Barreira:2018jgd}.
    
      Note that it is important to account for non-Gaussian contributions coupling different scales in order to address the relevance of the parameter dependence of the covariance matrix. As discussed in Section \ref{ssec:theory.single_map}, for a single Gaussian map, the relative bias between the PI and PD likelihoods drops like $\sim1/\ell$ for a single multipole order (see Eq. \ref{eq:bias_single}), and therefore as $\sim1/\ell_{\rm max}^2$ for a maximum multipole $\ell_{\rm max}$, becoming negligible for a sufficiently large number of modes. This will in general also be true for an arbitrary number of maps, since the same arguments hold for each of the independent eigenmaps that diagonalize $C^{ij}_\ell$. Non-Gaussian contributions to $\covar$ will couple different scales, effectively reducing the number of independent modes, and therefore may enhance the impact of the parameter-dependent covariance.
    
      The Gaussian contribution to the covariance matrix can be estimated as \cite{1995PhRvD..52.4307K}
      \begin{equation}\label{eq:covar_G}
        \left(\covar_{\rm G}\right)^{ij,\ell}_{mn,\ell'}=\delta_{\ell\ell'}\,\frac{C^{im}_\ell C^{jn}_\ell+C^{in}_\ell C^{jm}_\ell}{f_{\rm sky}(2\ell+1)}.
      \end{equation}
      Note that we account for an incomplete sky coverage by scaling the full-sky covariance by $1/f_{\rm sky}$ to account for the reduced number of available modes. This is not correct in detail, since measurements on a cut sky induce mode correlations, but it is a good enough approximation for forecasting \cite{2011MNRAS.414..329C}. It is also important to note that the power spectra entering Eq. \ref{eq:covar_G} must contain both signal and noise contributions. The former are given by Eq. \ref{eq:pspec_signal}, while the latter are
      \begin{figure*}
        \centering
        \includegraphics[width = 0.49\textwidth]{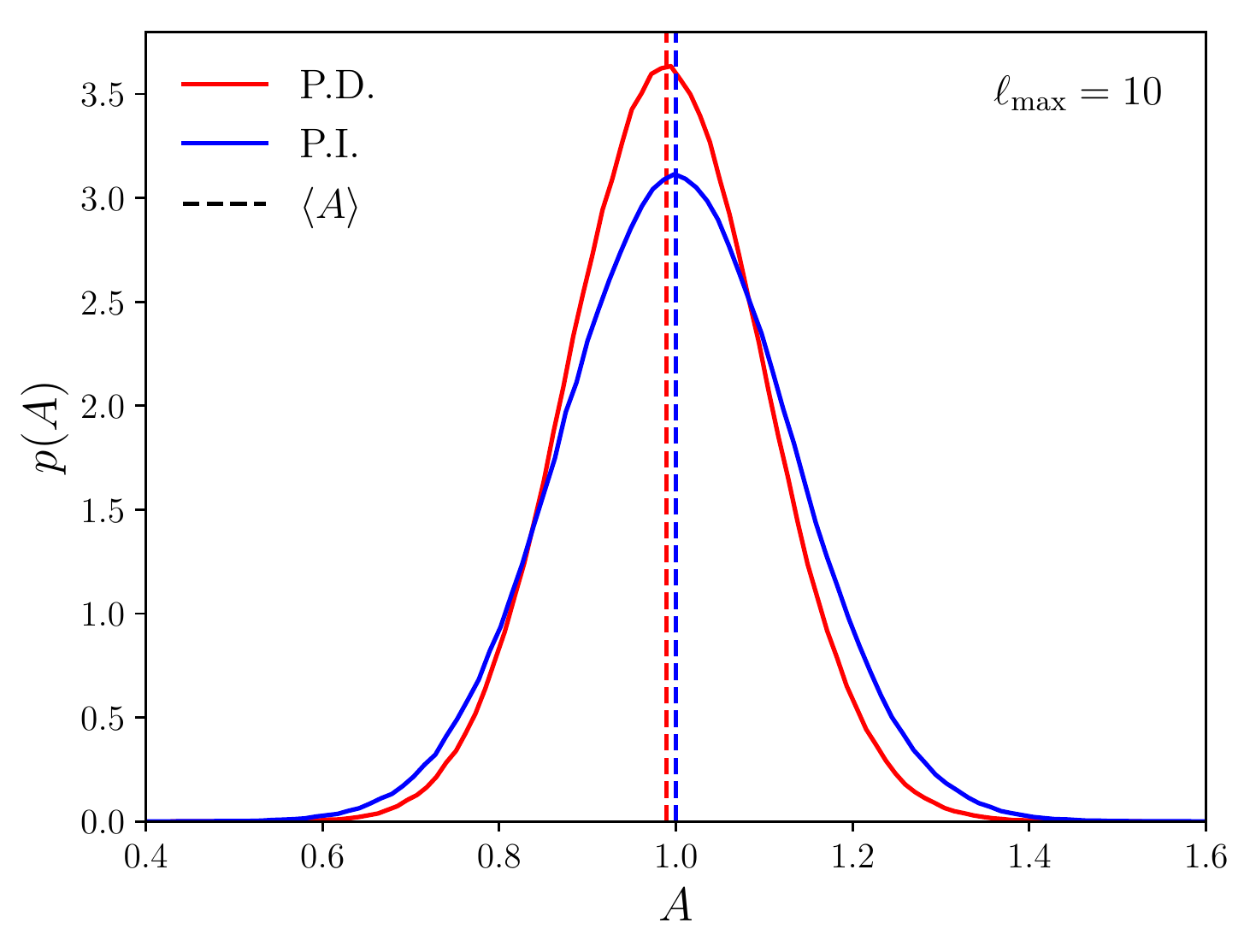}
        \includegraphics[width = 0.49\textwidth]{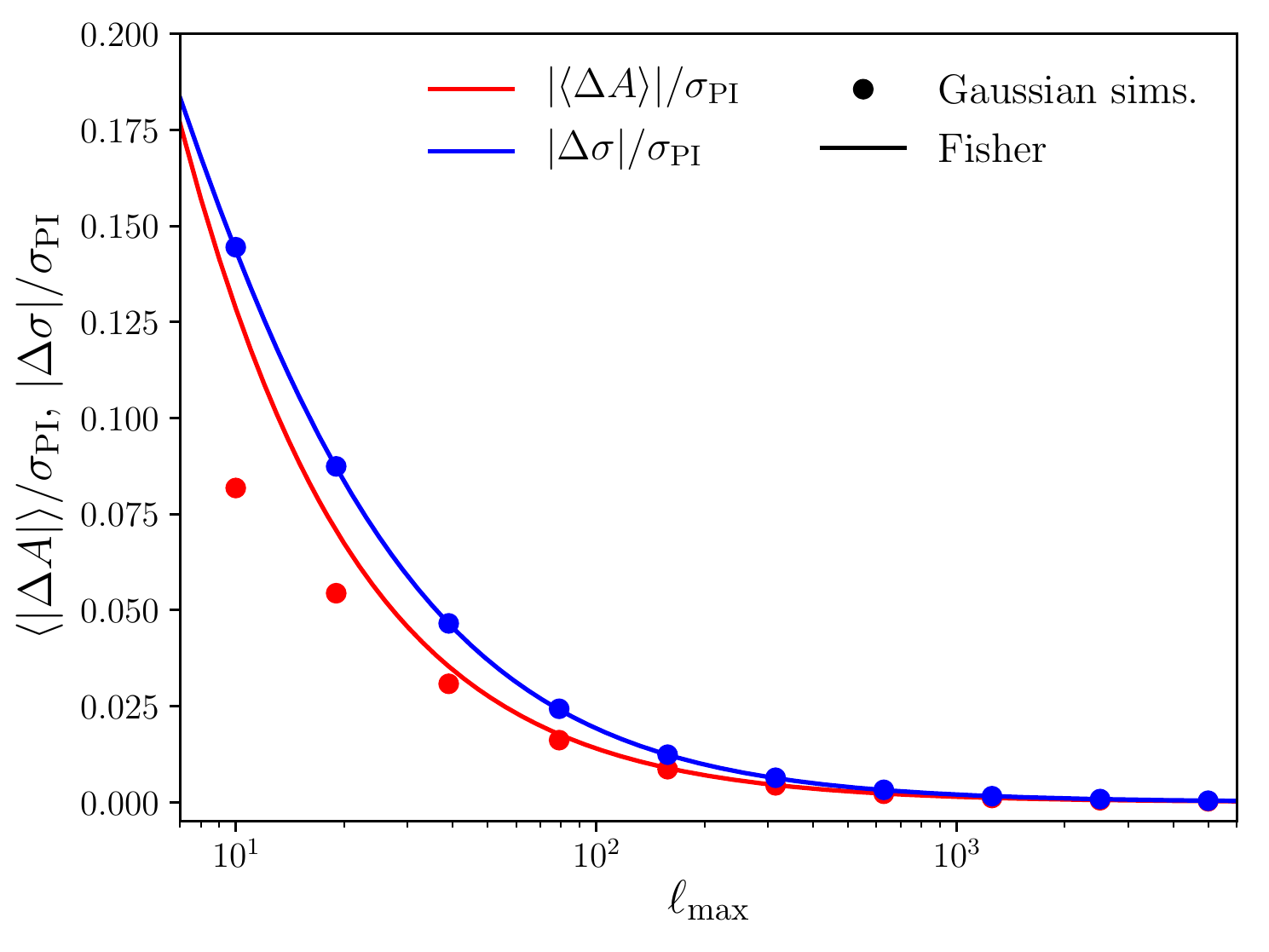}
        \caption{{\sl Left:} distribution of maximum-likelihood estimates of an overall power spectrum amplitude parameter in the case of parameter-dependent (red) and parameter-independent (blue) covariances. The parameter dependence of the covariance matrix causes a downwards relative bias with respect to the parameter-independent case, as well as a mild shrinkage of the parameter uncertainties. Results are shown for a small maximum multipole ($\ell_{\rm max}=10$ in order to highlight the effects of the parameter dependence. {\sl Right:} relative bias (red) and shift in uncertainties (blue) due to the parameter dependence of the covariance matrices as a function of the maximum multipole $\ell_{\rm max}$ (note that both quantities are normalized by the parameter-independent error bars). The red circles show the values found by directly evaluating the exact likelihood for $10^6$ simulations, while the solid lines show the values found in our Fisher matrix approximation. The Fisher approach is able to reproduce the exact result with good accuracy for both quantities, slightly over-estimating the bias for small $\ell_{\rm max}$.}\label{fig:simple}
      \end{figure*}
      \begin{equation}\label{eq:pspec_noise}
        N^{\delta,ij}_\ell=\frac{1}{n_g^i}\delta_{ij},\hspace{12pt}N^{\gamma,ij}_\ell=\frac{\sigma_\gamma^2}{n_g^i}\delta_{ij},
      \end{equation}
      where $n_g^i$ is the mean angular number density of objects in the $i$-th redshift bin, and $\sigma_\gamma=0.28$ is the intrinsic ellipticity dispersion per component.

      We compute the SSC contribution to the covariance matrix as described in appendix \ref{app:SSC}. However, ignoring this contribution for the moment, we can use the simple dependence of the Gaussian covariance with $f_{\rm sky}$ to study the importance of a parameter-dependent covariance matrix as a function survey properties. Simple inspection of Eq. \ref{eq:bias_general} shows that, since $\mathcal{F}_{\rm PI}\propto f_{\rm sky}$, the parameter bias scales like $\propto f_{\rm sky}^{-1}$. On the other hand, the effect on parameter uncertainties is
      \begin{align}
        \sigma_\mu&=\sqrt{\left(\mathcal{F}^{-1}\right)_{\mu\mu}}\\
                  &=\sqrt{\left(\mathcal{F}_{\rm PI}+\Delta\mathcal{F}\right)^{-1}_{\mu\mu}}\\
                  &\simeq\sigma_\mu^{\rm PI}-\frac{1}{2}\frac{\left(\mathcal{F}^{-1}_{\rm PI}\,\Delta\mathcal{F}\,\mathcal{F}^{-1}_{\rm PI}\right)_{\mu\mu}}{\sigma_\mu^{\rm PI}},
      \end{align}
      where $\Delta\mathcal{F}$ is the second term in Eq. \ref{eq:fisher_PD}, and in the last line we have expanded to first order in this parameter. Since $\Delta\mathcal{F}$ does not scale with $f_{\rm sky}$, the correction to the parameter errors (given by the second term above) is $\propto f_{\rm sky}^{-3/2}$. Thus, in general \emph{the relevance of a parameter-dependent covariance matrix will decrease with the surveyed area}. This results holds also in the presence of non-Gaussian contributions to the covariance. The SSC term arises from the non-linear coupling induced between different modes by the presence of super-survey, long-wavelength modes. Its impact will therefore increase for smaller sky areas, and therefore we can expect it to induce a slightly steeper dependence on $f_{\rm sky}$ than the purely Gaussian case.
      
      The expected behaviour with the smallest scale included in the analysis $\ell_{\rm max}$ is also similar: increasing the number of modes reduces the relative impact of the parameter-dependent covariance. The non-Gaussian terms can somewhat modify this behaviour, due to mode coupling. However, the relative impact of the SSC contribution will also decrease for larger survey areas, or for more noise-dominated datasets.

    \subsection{Analytic example: the power spectrum amplitude}\label{ssec:theory.analytic}
      Before we study the importance of the parameter dependence in the covariance matrix in detail for the case of tomographic large-scale structure datasets, let us examine a particularly simpler scenario: a single sky map, and a single parameter quantifying the overall amplitude of the power spectrum (for instance, this would correspond to the case of $A_s$ or $\sigma_8$ for linear power spectra). In this case, we can calculate the impact of the parameter-dependent covariance exactly, and therefore it allows us to quantify the validity of the Fisher approximations derived in the previous section. Let us label the amplitude parameter $A$, such that our model for the measured angular power spectrum is related to the a fiducial one as
      \begin{equation}
        C_\ell = A C^f_\ell.
      \end{equation}
      We will assume a fiducial value of $A=1$, and consider a purely Gaussian covariance with no noise
      \begin{equation}
        \covar_{\ell\ell'} = \frac{(A\,C^f_\ell)^2}{n_\ell}\,\delta_{\ell\ell'},
      \end{equation}
      where $n_\ell=\ell+1/2$.
      
      On the one hand, the parameter-dependent and independent likelihoods are (up to irrelevant constants) given by 
      \begin{align}
        G_{\rm PD}(A)&=\sum_{\ell=0}^{\ell_{\rm max}}n_\ell\left(\frac{r_\ell}{A}-1\right)^2+2(\ell_{\rm max}+1)\ln(A)\\
        G_{\rm PI}(A)&=\sum_{\ell=0}^{\ell_{\rm max}}n_\ell\left(r_\ell-A\right)^2,
      \end{align}
      where $r_\ell=C^d_\ell/C^f_\ell$, and $C^d_\ell$ is the measured power spectrum. The maximum-likelihood solutions for $A$ for a given realisation of the data then are
      \begin{align}
        \hat{A}_{\rm PD}&=\frac{S_1}{2(\ell_{\rm max}+1)}\left[\sqrt{1+4\frac{(\ell_{\rm max}+1)S_2}{(S_1)^2}}-1\right]\\
        \hat{A}_{\rm PI}&=\frac{S_1}{S_0}
      \end{align}
      where $S_n\equiv\sum_\ell n_\ell r_\ell^n$. On the other hand, our Fisher predictions for the parameter bias and the parameter-independent and parameter-dependent errors (Eqs. \ref{eq:fisher_bias} and \ref{eq:fisher_PD}) are
      \begin{align}\label{eq:bias_simple}
        \langle\Delta A\rangle&=-(S_0)^{-1}=-\frac{2}{(\ell_{\rm max}+1)^2},\\\label{eq:sigmaPI_simple}
        \sigma_{\rm PI}(A)&= (S_0)^{-1/2}=\frac{\sqrt{2}}{\ell_{\rm max}+1},\\\label{eq:sigmaPD_simple}
        \sigma_{\rm PD}(A)&= \sigma_{\rm PI}(A)\left[1+\frac{4}{(\ell_{\rm max}+1)}\right]^{-1/2}
      \end{align}
      where $\ell_{\rm max}$ is the maximum multipole included in the analysis.
      
      To validate these results, we generate $10^6$ random Gaussian realisations of $r_\ell$ with standard deviation $\sigma_\ell=n_\ell^{-1}$, and compute $\hat{A}_{\rm PD}$ and $\hat{A}_{\rm PI}$ for each of them. We then evaluate the mean and standard deviation of both quantities and compare them with the approximate results in Eqs. \ref{eq:bias_simple}, \ref{eq:sigmaPI_simple} and \ref{eq:sigmaPD_simple}. The results of this validation are shown in Figure \ref{fig:simple}. The left panel shows an example of the distributions of $\hat{A}_{\rm PI}$ (blue) and $\hat{A}_{\rm PD}$ (red) for the case of $\ell_{\rm max}=10$. As discussed above, the small number of modes in this case highlights the relevance of the parameter-dependent covariance, which produces a noticeable relative bias on $A$ and a decrease in its uncertainty \cite{Hamimeche:2008ai}. The right panel then shows the relative parameter bias and shift in uncertainties normalised by $\sigma_{\rm PI}$ as a function of $\ell_{\rm max}$ for the simulated likelihoods (circles) and for our Fisher predictions (solid lines). The Fisher approximation is remarkably good, and remains accurate at the $10\%$ level even for low values of $\ell_{\rm max}\sim30$.

  \section{Forecasts}\label{sec:forecasts}
    We apply the results discussed in the previous section to the case of imaging surveys targeting a joint analysis of galaxy clustering and weak lensing. We start by describing the assumptions we use to quantify the expected signal and noise of these surveys and then present our results regarding the relevance of the parameter dependence of the covariance matrix.
    
    \subsection{Survey specifications}\label{ssec:forecasts.surveys}
      \begin{figure}
        \centering 
        \includegraphics[width = 0.47\textwidth]{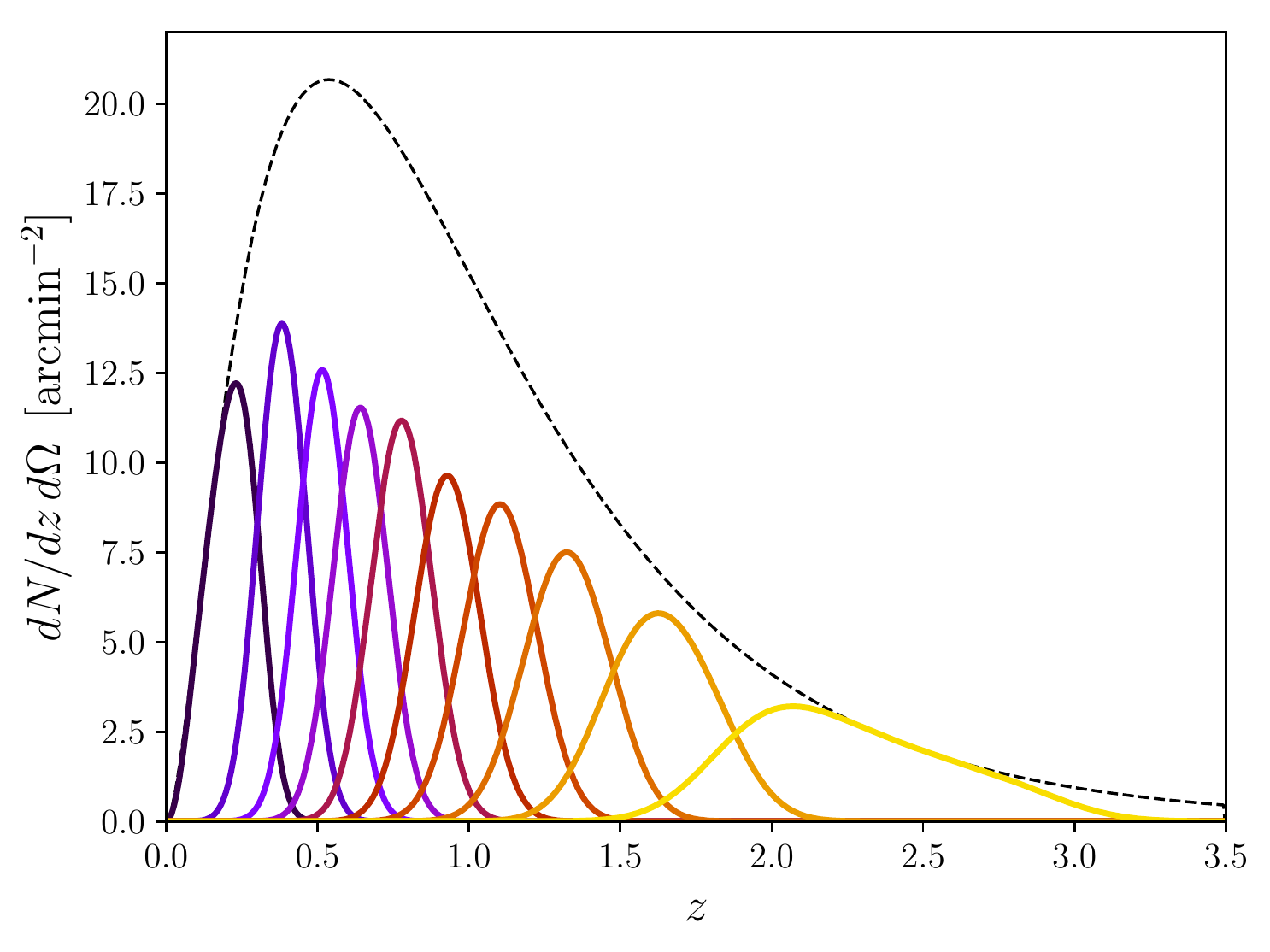}
        \caption{Redshift distribution for the overall sample assumed in this analysis (dashed black) and for each of the 10 redshift bins (solid coloured lines). For simplicity we assume the same redshift distribution for both lenses and sources.} \label{fig:dndz}
      \end{figure}
      We assume an LSST-like survey. For simplicity we assume the same redshift distribution for the clustering and lensing samples with a redshift dependence \cite{2018arXiv180901669T}
      \begin{equation}
        \frac{dN}{dz}\propto z^2\exp\left[-\left(\frac{z}{z_0}\right)^\alpha\right],\hspace{12pt}
        (z_0,\alpha)=(0.12,0.7),
      \end{equation}
      and a total number density $n_g=27\,{\rm arcmin}^{-2}$. We split the total sample into 10 photometric redshift bins with approximately equal number of galaxies in each bin. The redshift distributions for each bin are calculated assuming Gaussian photometric redshift uncertainties with a standard deviation
      $\sigma_z=0.05\,(1+z)$. These are shown in Figure \ref{fig:dndz}.

      We produce forecasts for a set of 4 cosmological parameters: the fractional matter density $\Omega_M$, the matter power spectrum amplitude $\sigma_8$ and equation of state parameters $w_0$ and $w_a$. The forecasts presented below are marginalized over 6 nuisance parameters corresponding to the values of the linear galaxy bias in 6 redshift nodes $z=(0.25,0.75,1.25,1.85,2.60,3.25)$. We assume a linear dependence with redshift between these nodes, and our fiducial bias values correspond to $b(z)=1+0.84\,z$ \cite{2004ApJ...601....1W}. Since the aim of this work is to explore the impact of parameter-dependent covariances, and not to produce forecast of the expected cosmological constraints, we do not consider any other sources of systematic uncertainty, such as intrinsic alignments, multiplicative shape biases or photometric redshift systematics.
      
      Finally, our fiducial forecasts assume a sky fraction $f_{\rm sky}=0.4$, as expected for LSST, and a constant scale cut $\ell_{\rm max}=3000$ for all redshift bins. A more realistic choice of scale cuts would be motivated by modelling uncertainties, by removing all scales smaller than the physical scale of non-linearities $k_{\rm NL}$, necessarily in a redshift-dependent way. By not removing these scales at low redshifts ($\ell_{\rm NL}(z=0.5)\simeq k_{\rm NL}\chi(z=0.5)\simeq400$ for $k_{\rm NL}=0.3\,h\,{\rm Mpc}^{-1}$), our fiducial choice emphasizes the role of super-sample covariance, potentially highlighting the effects of parameter dependence in the covariance. At the same time, we have seen that these effects become less relevants as we increase the number of independent modes, and therefore we will also present results as a function of $\ell_{\rm max}$ and $f_{\rm sky}$.
      
    \subsection{Results}\label{ssec:forecasts.results}
      \begin{figure*}
        \centering 
        \includegraphics[scale = 0.45]{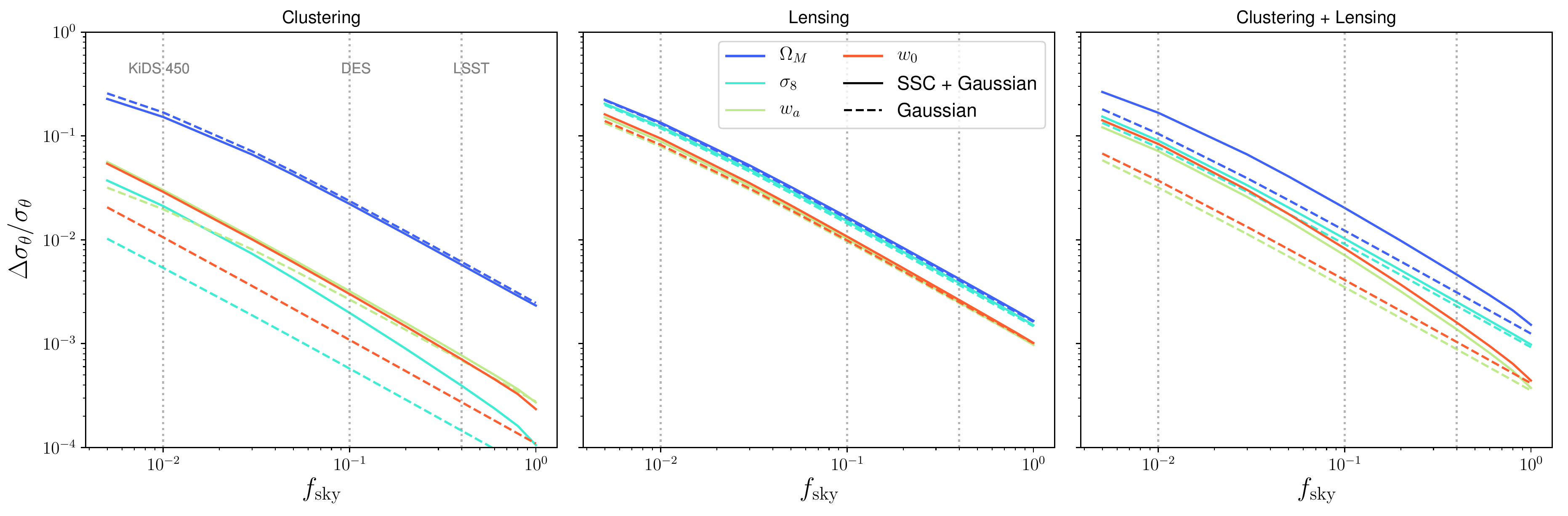}
        \caption{Relative difference in the uncertainties on late-time cosmological parameters due to the parameter dependence of the covariance matrix as a function of survey area for a fiducial $\ell_{\rm max}=3000$.  From left to right, we show results for galaxy clustering, weak lensing and for the combination of both probes. Results are shown for Gaussian covariance matrices (dashed lines) and including the super-sample covariance (solid lines). As argued in Section \ref{sec:theory}, the importance of the parameter dependence grows towards smaller $f_{\rm sky}$, as the number of modes available decreases. Nevertheless, the effect is never larger than $\sim20\%$ of the statistical errors for the smallest sky fractions ($f_{\rm sky}=0.005$), and is below $1\%$ for LSST-like areas ($f_{\rm sky}\simeq0.4$). } \label{fig:rel_error_fsky}
      \end{figure*}
      
      We compute the Fisher matrices for parameter-dependent and independent covariances as well as the bias due to the parameter dependence as described in Section \ref{ssec:theory.general}, using Eqs. \ref{eq:fisher_PI}, \ref{eq:fisher_PD} and \ref{eq:fisher_bias}. Here ${\bf t}$ is the vector of all possible cross-power spectra between two tracers ($\delta_g$ or $\gamma$) in any pair of redshift bins, calculated using Eq. \ref{eq:multinorm} (with noise power spectra given by Eq. \ref{eq:pspec_noise}), and $\covar$ is the covariance matrix of this data vector, calculated as in Eq. \ref{eq:covar_lss}. To compute all angular power spectra we use the Core Cosmology Library\footnote{\url{https://github.com/LSSTDESC/CCL}} \cite{ccl}, which we also modified to provide estimates of the super-sample covariance as described in Appendix \ref{app:SSC}. Note that, we do not compute power spectra and covariance matrices for all integer values of $\ell$. Instead, we use 15 logarithmically-spaced bandpowers between $\ell=20$ and $\ell=3000$.
      
      We report our results in terms of the relative parameter biases and the relative error differences
      \begin{equation}
        \frac{\Delta\theta_\mu}{\sigma_{{\rm PI},\mu}},\hspace{12pt}
        \frac{\Delta\sigma_\mu}{\sigma_{{\rm PI},\mu}}\equiv\frac{\sigma_{{\rm PI},\mu}-\sigma_{{\rm PD},\mu}}{\sigma_{{\rm PI},\mu}},
      \end{equation}
      where $\sigma_{\rm PI}$ and $\sigma_{\rm PD}$ are computed from the inverse of the corresponding Fisher matrices. Results are reported for the 4 cosmological parameters $(\Omega_M,\sigma_8,w_0,w_a)$. Since the aim of this paper is to study the relevance of the parameter dependence in the covariance matrix, and not to produce cosmological forecasts for future surveys, we do not report absolute errors on these parameters.
      \begin{figure*}
        \centering 
        \includegraphics[scale = 0.45]{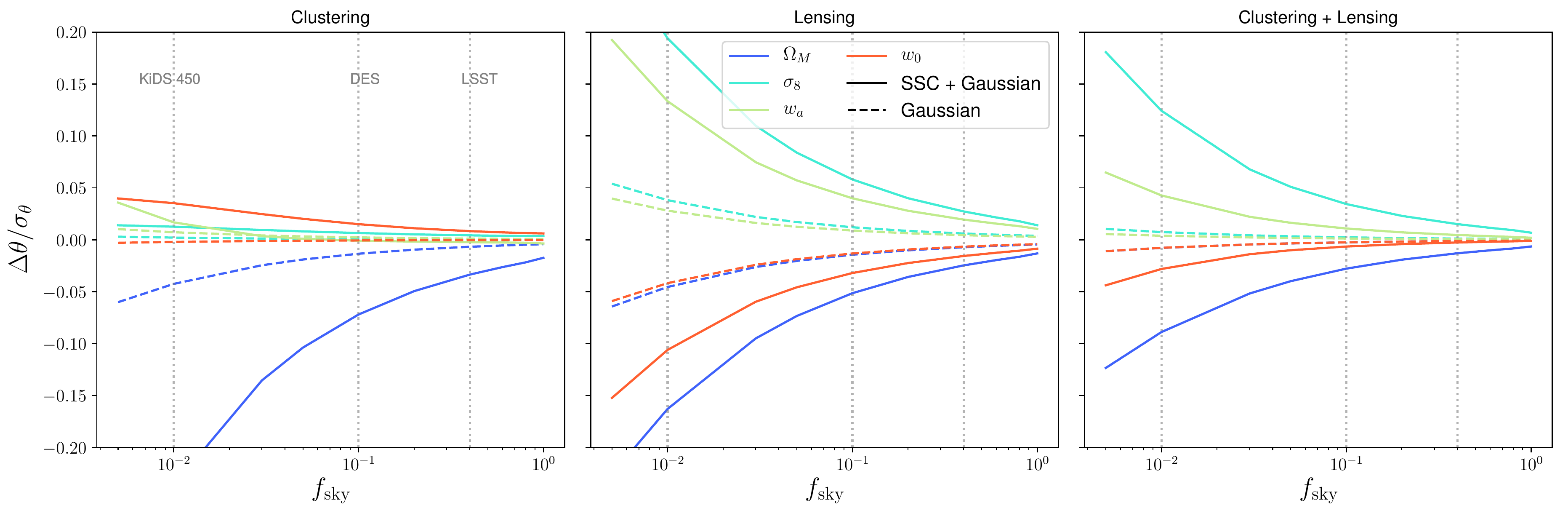}
        \caption{Same as Fig. \ref{fig:rel_error_fsky} for the relative parameter bias. The effects of a parameter-dependent covariance are always smaller than $30\%$ of the statistical uncertainties for the smallest sky areas ($\sim0.5\%$ of the sky) and become percent-level for LSST-like areas ($f_{\rm sky}\simeq0.4$).} \label{fig:rel_bias_fsky}
      \end{figure*}
      \begin{figure*}
        \centering 
        \includegraphics[width = 0.47\textwidth]{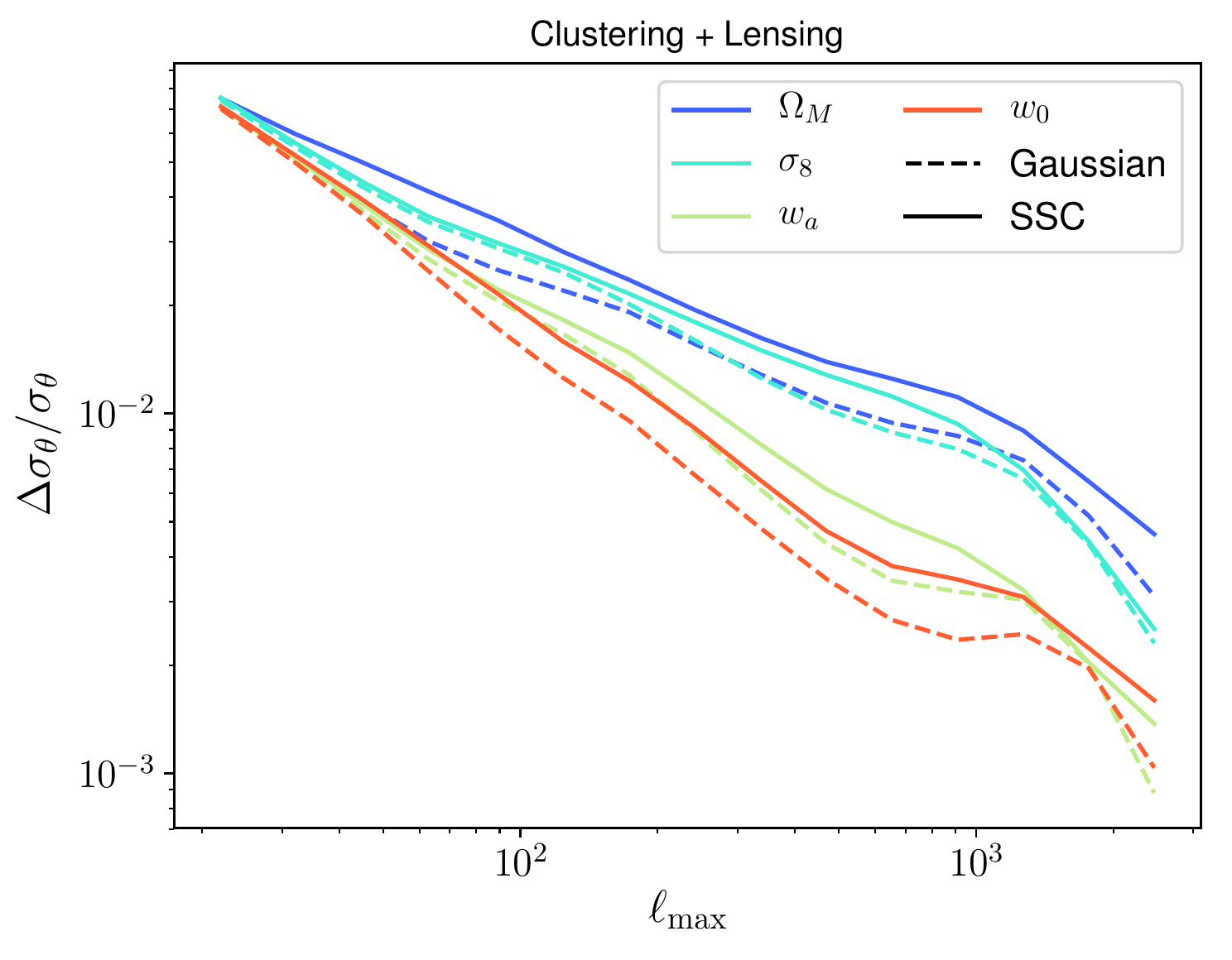}
        \includegraphics[width = 0.47\textwidth]{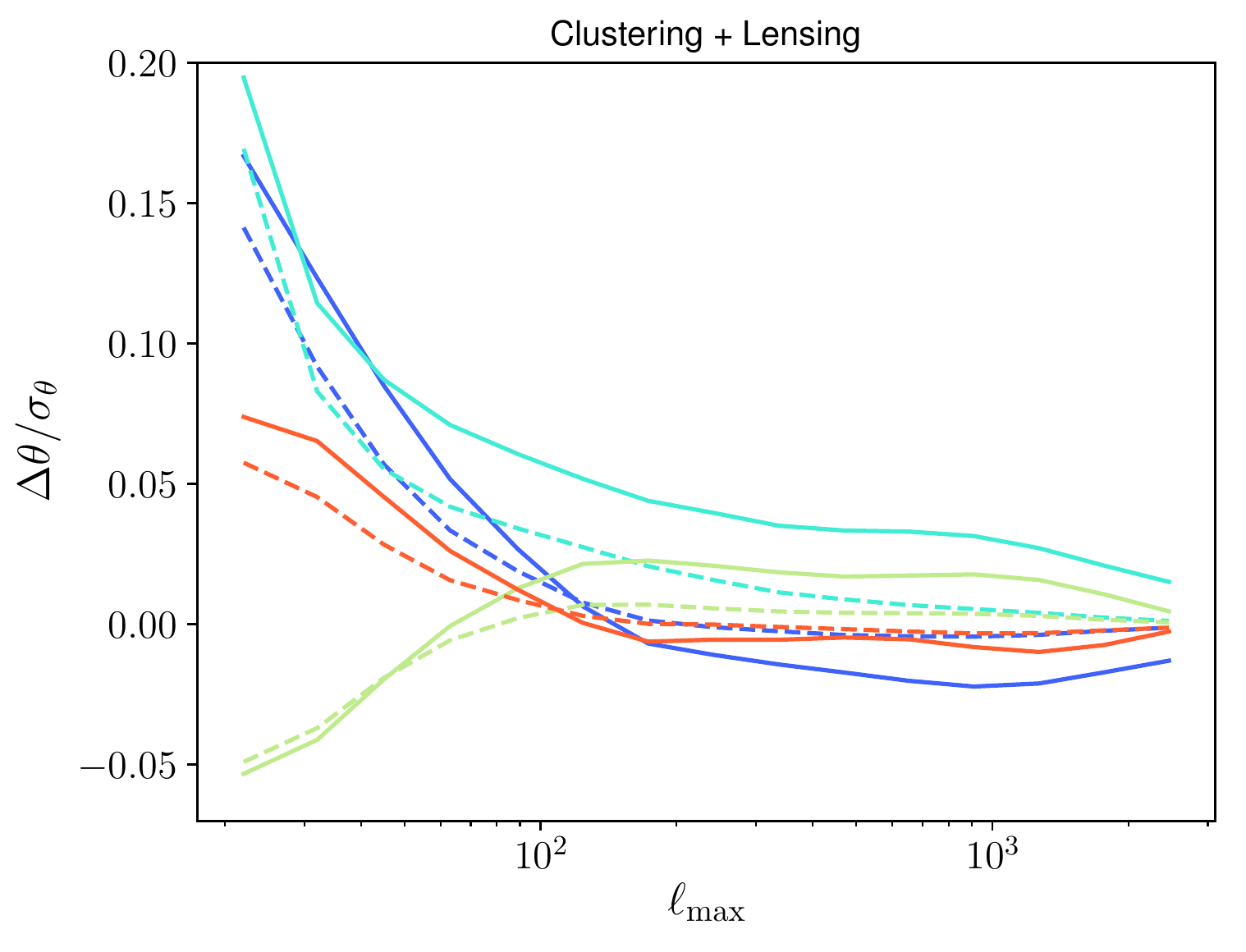}
        \caption{{\sl Left:} relative error difference due to a parameter-dependent covariance for the four cosmological parameters studied here as a function of the maximum multipole $\ell_{\rm max}$ included in the analysis for a fiducial sky fraction $f_{\rm sky}=0.4$. {\sl Right:} same as the left panel for the relative parameter bias. As expected, the impact of the parameter dependence grows as we reduce the number of independent modes used in the analysis. In all cases, the effect is smaller than $\sim20\%$ of the statistical uncertainties, even for $\ell_{\rm max}\sim20$, and becomes percent-level for realistic values ($\ell_{\rm max}\sim1000$).} \label{fig:pd_lmax}
      \end{figure*}
      
      Figure \ref{fig:rel_error_fsky} shows the relative difference in the statistical uncertainties as a function of sky area for a fiducial $\ell_{\rm max}=3000$. Results are shown for galaxy clustering, weak lensing and for the combination of both. We also show the impact of the super-sample contribution to the covariance matrix by displaying results with (solid) and without it (dashed). In all cases, the information contained in the covariance matrix is very small, and would only lead to a $\sim20\%$ suppression of the uncertainties for the smallest sky areas ($\sim0.5\%$ of the sky), corresponding to e.g. CFHTLens \cite{2013MNRAS.433.2545E} or the first data release of HSC \cite{2018arXiv180909148H}. The effect becomes even less important for larger sky areas, as more independent modes become available and concentrate most of the information on the power spectrum. For LSST-like areas ($f_{\rm sky}\simeq0.4$) the effect is, at most, of the order of $1\%$ of the statistical uncertainties.
      
      The same results are shown in Figure \ref{fig:rel_bias_fsky} for the relative parameter biases, and similar conclusions hold. The effects are always smaller than $\sim0.3\sigma$ on small survey areas, and get suppressed to a percent fraction of the statistical uncertainties for larger areas. While these effects are small, it is interesting to note that they are enhanced by considering a more accurate model of the covariance matrix that includes the SSC term. Thus we see once again the effect of the super-sample covariance inducing a statistical coupling between modes which then reduces the effective number of independent degrees of freedom and and enhances the relative information content of the covariance matrix.

      For a fixed LSST-like sky fraction ($f_{\rm sky}=0.4$), Figure \ref{fig:pd_lmax} shows the impact of the parameter dependence of the covariance matrix as a function of the maximum multipole $\ell_{\rm max}$ included in the analysis. Results are shown for the error difference (left panel) and biases (right panel), and we also show the impact of the super-sample covariance term. Again, as expected based on our discussion in Section \ref{sec:theory}, the information content of the covariance matrix decreases as more independent modes are included in the analysis. The impact, both on the uncertainties and on the parameter biases, is smaller than $\sim20\%$ of the overall statistical error budget (and this only for $\ell_{\rm max}=20$). For more realistic scale cuts ($\ell_{\rm max}\sim1000$), the effect is suppressed to the level of ${\cal O}(1\%)$. 

  \section{Summary}\label{sec:summary}
    In the era of precision cosmology it is becoming increasingly important to understand the systematic and statistical errors that occur during parameter estimation. In this paper we have analysed the effect of using a likelihood with a parameter dependent covariance matrix on the inference of cosmological parameters. The computation of covariance matrices for large-scale structure surveys is a numerically complex problem for which multiple approaches have been proposed in the literature. Therefore, assessing the need to estimate the covariance at every point of a given likelihood evaluation, rather than estimating it only once for a set of parameters sufficiently close to the maximum likelihood, is important for future surveys, where covariance estimation will become more computationally demanding.
    
    We have focused our analysis on the two main aspects of parameter inference: final statistical uncertainties and parameter biases. For multivariate Gaussian likelihoods, we have quantified the impact on the statistical errors using a standard Fisher matrix approach that accounts for the parameter dependence of both the mean and covariance (Eq. \ref{eq:fisher_PD}). We have also derived an expression to estimate the expected parameter bias by expanding the likelihood around its maximum. The resulting expression (Eq. \ref{eq:fisher_bias}) is easy to calculate and has not been presented before to our knowledge. We have evaluated the accuracy of this approximation by comparing its predictions with the analytical solutions available for a simplified case involving a single amplitude parameter and a single sky map (Section \ref{ssec:theory.analytic}). This exercise shows that our approximate estimates are indeed accurate as long as the true parameter bias is small, and that, if anything, our approximations will slightly overestimate this bias. The methods used here are therefore perfectly applicable to the case we study, and a more computationally expensive approach involving a full evaluation of the likelihood through Monte-Carlo methods is unlikely to yield different results.
    
    We have then evaluated the parameter shifts and error differences for the particular case of a large-scale structure experiment targeting the joint measurement of galaxy clustering and cosmic shear. The Fisher information for the parameter dependent covariance matrix, in contrast to the parameter independent case, does not increase with the survey area. This can be seen in Eq \ref{eq:fisher_PD} where all factors of $f_{\rm sky}$ in the covariance matrix cancel out for the second term. The associated parameter bias, on the other hand, is roughly inversely proportional to $f_{\rm sky}$ (see Section \ref{ssec:theory.lss}). This means there will always be a regime in which the parameter dependence becomes important, the question is whether it will be important for any practical value of $f_{\rm sky}$. Our study of simple single-map cases (Sections \ref{ssec:theory.single_map} and \ref{ssec:theory.analytic}) have shown that the impact of a parameter-dependent covariance decreases with multipole order (e.g. Eqs. \ref{eq:bias_simple} and \ref{eq:sigmaPD_simple}), and therefore the generic message is that \emph{the relative information content in a parameter-dependent covariance decreases as more independent modes are included in the survey}. To account for mode-coupling induced by non-linear evolution of the matter overdensities, we have included the super-sample contribution to the covariance matrix, which has been determined to be the most relevant contribution in the range of scales considered here \cite{Mohammed:2016sre}. We note that we have not considered additional nuisance parameters such as shifts in the photometric redshifts, the impact of baryons or intrinsic alignments. We have also only taken into account the super-sample covariance contribution to the Gaussian covariance matrix, neglecting all other parts of the connected trispectrum. While this has been determined to be sufficiently accurate for lensing observables \cite{Barreira:2018jgd} large-scale structure studies exploring the deeply non-linear regime will require a more careful treatment. Nevertheless, we don not expect any of these effects to change our results significantly. Furthermore it is conceivable that in situations where the parameter space is extended, to include more exotic models of dark energy for example, the likelihood may become Non-Gaussian and then the parameter dependence of the covariance may need to be taken into account. 
    
	Our findings, summarised in Figures \ref{fig:rel_error_fsky}, \ref{fig:rel_bias_fsky} and \ref{fig:pd_lmax} show that, for any current and future surveys, the parameter dependence of the covariance matrix can be safely ignored, since it only leads to changes in the statistical errors and maximum-likelihood parameters that are $\lesssim1\%$ of the statistical uncertainties. For surveys targeting very small sky fractions ($f_{\rm sky}<1\%$) or very large scales ($\ell\lesssim20$), the impact of the parameter dependence becomes more important, but the impact is at most at the $\sim0.2\sigma$ level, both in biases and uncertainties. Note that the these effects are sufficiently small that it strongly justifies the use of the Fisher formalism to undertake this estimate, but a more thorough analysis of these effects in cases where information content of the covariance is significant would require the use of a full likelihood exploration. Since we can expect systematic and numerical uncertainties to be at least as important, it is safe to say that \emph{the parameter dependence of the covariance matrix can be ignored in all practical cases}. 
    
  \acknowledgments
    We thank Harry Desmond, Tim Eifler, Alan Heavens, Elisabeth Krause and Elena Sellentin for useful comments and discussions. DK and PGF acknowledge support from the ERC. DA acknowledges support from the Beecroft trust and from the Science and Technology Facilities Council (STFC) through an Ernest Rutherford Fellowship, grant reference ST/P004474/1. PGF acknowledges support from the Beecroft trust and STFC.

  \bibliographystyle{apsrev4-1}
  \bibliography{main}

\begin{thebibliography}{53}%
\makeatletter
\providecommand \@ifxundefined [1]{%
 \@ifx{#1\undefined}
}%
\providecommand \@ifnum [1]{%
 \ifnum #1\expandafter \@firstoftwo
 \else \expandafter \@secondoftwo
 \fi
}%
\providecommand \@ifx [1]{%
 \ifx #1\expandafter \@firstoftwo
 \else \expandafter \@secondoftwo
 \fi
}%
\providecommand \natexlab [1]{#1}%
\providecommand \enquote  [1]{``#1''}%
\providecommand \bibnamefont  [1]{#1}%
\providecommand \bibfnamefont [1]{#1}%
\providecommand \citenamefont [1]{#1}%
\providecommand \href@noop [0]{\@secondoftwo}%
\providecommand \href [0]{\begingroup \@sanitize@url \@href}%
\providecommand \@href[1]{\@@startlink{#1}\@@href}%
\providecommand \@@href[1]{\endgroup#1\@@endlink}%
\providecommand \@sanitize@url [0]{\catcode `\\12\catcode `\$12\catcode
  `\&12\catcode `\#12\catcode `\^12\catcode `\_12\catcode `\%12\relax}%
\providecommand \@@startlink[1]{}%
\providecommand \@@endlink[0]{}%
\providecommand \url  [0]{\begingroup\@sanitize@url \@url }%
\providecommand \@url [1]{\endgroup\@href {#1}{\urlprefix }}%
\providecommand \urlprefix  [0]{URL }%
\providecommand \Eprint [0]{\href }%
\providecommand \doibase [0]{http://dx.doi.org/}%
\providecommand \selectlanguage [0]{\@gobble}%
\providecommand \bibinfo  [0]{\@secondoftwo}%
\providecommand \bibfield  [0]{\@secondoftwo}%
\providecommand \translation [1]{[#1]}%
\providecommand \BibitemOpen [0]{}%
\providecommand \bibitemStop [0]{}%
\providecommand \bibitemNoStop [0]{.\EOS\space}%
\providecommand \EOS [0]{\spacefactor3000\relax}%
\providecommand \BibitemShut  [1]{\csname bibitem#1\endcsname}%
\let\auto@bib@innerbib\@empty
%</preamble>
\bibitem [{\citenamefont {{LSST Science Collaboration}}\ \emph
  {et~al.}(2009)\citenamefont {{LSST Science Collaboration}} \emph
  {et~al.}}]{2009arXiv0912.0201L}%
  \BibitemOpen
  \bibfield  {author} {\bibinfo {author} {\bibnamefont {{LSST Science
  Collaboration}}} \emph {et~al.},\ }\href@noop {} {\bibfield  {journal}
  {\bibinfo  {journal} {ArXiv e-prints}\ } (\bibinfo {year} {2009})},\ \Eprint
  {http://arxiv.org/abs/0912.0201} {arXiv:0912.0201 [astro-ph.IM]} \BibitemShut
  {NoStop}%
\bibitem [{\citenamefont {{Laureijs}}\ \emph {et~al.}(2011)\citenamefont
  {{Laureijs}} \emph {et~al.}}]{2011arXiv1110.3193L}%
  \BibitemOpen
  \bibfield  {author} {\bibinfo {author} {\bibfnamefont {R.}~\bibnamefont
  {{Laureijs}}} \emph {et~al.},\ }\href@noop {} {\bibfield  {journal} {\bibinfo
   {journal} {ArXiv e-prints}\ } (\bibinfo {year} {2011})},\ \Eprint
  {http://arxiv.org/abs/1110.3193} {arXiv:1110.3193 [astro-ph.CO]} \BibitemShut
  {NoStop}%
\bibitem [{\citenamefont {{Spergel}}\ \emph {et~al.}(2013)\citenamefont
  {{Spergel}} \emph {et~al.}}]{2013arXiv1305.5422S}%
  \BibitemOpen
  \bibfield  {author} {\bibinfo {author} {\bibfnamefont {D.}~\bibnamefont
  {{Spergel}}} \emph {et~al.},\ }\href@noop {} {\bibfield  {journal} {\bibinfo
  {journal} {ArXiv e-prints}\ } (\bibinfo {year} {2013})},\ \Eprint
  {http://arxiv.org/abs/1305.5422} {arXiv:1305.5422 [astro-ph.IM]} \BibitemShut
  {NoStop}%
\bibitem [{\citenamefont {{DES Collaboration}}\ \emph
  {et~al.}(2017)\citenamefont {{DES Collaboration}} \emph
  {et~al.}}]{2017arXiv170801530D}%
  \BibitemOpen
  \bibfield  {author} {\bibinfo {author} {\bibnamefont {{DES Collaboration}}}
  \emph {et~al.},\ }\href@noop {} {\bibfield  {journal} {\bibinfo  {journal}
  {ArXiv e-prints}\ } (\bibinfo {year} {2017})},\ \Eprint
  {http://arxiv.org/abs/1708.01530} {arXiv:1708.01530} \BibitemShut {NoStop}%
\bibitem [{\citenamefont {{Joudaki}}\ \emph {et~al.}(2018)\citenamefont
  {{Joudaki}} \emph {et~al.}}]{2018MNRAS.474.4894J}%
  \BibitemOpen
  \bibfield  {author} {\bibinfo {author} {\bibfnamefont {S.}~\bibnamefont
  {{Joudaki}}} \emph {et~al.},\ }\href {\doibase 10.1093/mnras/stx2820}
  {\bibfield  {journal} {\bibinfo  {journal} {\mnras}\ }\textbf {\bibinfo
  {volume} {474}},\ \bibinfo {pages} {4894} (\bibinfo {year} {2018})},\ \Eprint
  {http://arxiv.org/abs/1707.06627} {arXiv:1707.06627} \BibitemShut {NoStop}%
\bibitem [{\citenamefont {{Hikage}}\ \emph {et~al.}(2018)\citenamefont
  {{Hikage}} \emph {et~al.}}]{2018arXiv180909148H}%
  \BibitemOpen
  \bibfield  {author} {\bibinfo {author} {\bibfnamefont {C.}~\bibnamefont
  {{Hikage}}} \emph {et~al.},\ }\href@noop {} {\bibfield  {journal} {\bibinfo
  {journal} {ArXiv e-prints}\ } (\bibinfo {year} {2018})},\ \Eprint
  {http://arxiv.org/abs/1809.09148} {arXiv:1809.09148} \BibitemShut {NoStop}%
\bibitem [{\citenamefont {Zentner}\ \emph {et~al.}(2013)\citenamefont
  {Zentner}, \citenamefont {Semboloni}, \citenamefont {Dodelson}, \citenamefont
  {Eifler}, \citenamefont {Krause},\ and\ \citenamefont
  {Hearin}}]{PhysRevD.87.043509}%
  \BibitemOpen
  \bibfield  {author} {\bibinfo {author} {\bibfnamefont {A.~R.}\ \bibnamefont
  {Zentner}}, \bibinfo {author} {\bibfnamefont {E.}~\bibnamefont {Semboloni}},
  \bibinfo {author} {\bibfnamefont {S.}~\bibnamefont {Dodelson}}, \bibinfo
  {author} {\bibfnamefont {T.}~\bibnamefont {Eifler}}, \bibinfo {author}
  {\bibfnamefont {E.}~\bibnamefont {Krause}}, \ and\ \bibinfo {author}
  {\bibfnamefont {A.~P.}\ \bibnamefont {Hearin}},\ }\href {\doibase
  10.1103/PhysRevD.87.043509} {\bibfield  {journal} {\bibinfo  {journal} {Phys.
  Rev. D}\ }\textbf {\bibinfo {volume} {87}},\ \bibinfo {pages} {043509}
  (\bibinfo {year} {2013})}\BibitemShut {NoStop}%
\bibitem [{\citenamefont {Rudd}\ \emph {et~al.}(2008)\citenamefont {Rudd},
  \citenamefont {Zentner},\ and\ \citenamefont
  {Kravtsov}}]{0004-637X-672-1-19}%
  \BibitemOpen
  \bibfield  {author} {\bibinfo {author} {\bibfnamefont {D.~H.}\ \bibnamefont
  {Rudd}}, \bibinfo {author} {\bibfnamefont {A.~R.}\ \bibnamefont {Zentner}}, \
  and\ \bibinfo {author} {\bibfnamefont {A.~V.}\ \bibnamefont {Kravtsov}},\
  }\href {http://stacks.iop.org/0004-637X/672/i=1/a=19} {\bibfield  {journal}
  {\bibinfo  {journal} {The Astrophysical Journal}\ }\textbf {\bibinfo {volume}
  {672}},\ \bibinfo {pages} {19} (\bibinfo {year} {2008})}\BibitemShut
  {NoStop}%
\bibitem [{\citenamefont {van~den Bosch}\ \emph {et~al.}(2013)\citenamefont
  {van~den Bosch}, \citenamefont {More}, \citenamefont {Cacciato},
  \citenamefont {Mo},\ and\ \citenamefont {Yang}}]{vdb}%
  \BibitemOpen
  \bibfield  {author} {\bibinfo {author} {\bibfnamefont {F.~C.}\ \bibnamefont
  {van~den Bosch}}, \bibinfo {author} {\bibfnamefont {S.}~\bibnamefont {More}},
  \bibinfo {author} {\bibfnamefont {M.}~\bibnamefont {Cacciato}}, \bibinfo
  {author} {\bibfnamefont {H.}~\bibnamefont {Mo}}, \ and\ \bibinfo {author}
  {\bibfnamefont {X.}~\bibnamefont {Yang}},\ }\href {\doibase
  10.1093/mnras/sts006} {\bibfield  {journal} {\bibinfo  {journal} {Monthly
  Notices of the Royal Astronomical Society}\ }\textbf {\bibinfo {volume}
  {430}},\ \bibinfo {pages} {725} (\bibinfo {year} {2013})}\BibitemShut
  {NoStop}%
\bibitem [{\citenamefont {{Schneider}}\ and\ \citenamefont
  {{Teyssier}}(2015)}]{2015JCAP...12..049S}%
  \BibitemOpen
  \bibfield  {author} {\bibinfo {author} {\bibfnamefont {A.}~\bibnamefont
  {{Schneider}}}\ and\ \bibinfo {author} {\bibfnamefont {R.}~\bibnamefont
  {{Teyssier}}},\ }\href {\doibase 10.1088/1475-7516/2015/12/049} {\bibfield
  {journal} {\bibinfo  {journal} {\jcap}\ }\textbf {\bibinfo {volume} {12}},\
  \bibinfo {eid} {049} (\bibinfo {year} {2015})},\ \Eprint
  {http://arxiv.org/abs/1510.06034} {arXiv:1510.06034} \BibitemShut {NoStop}%
\bibitem [{\citenamefont {{Chisari}}\ \emph {et~al.}(2018)\citenamefont
  {{Chisari}}, \citenamefont {{Richardson}}, \citenamefont {{Devriendt}},
  \citenamefont {{Dubois}}, \citenamefont {{Schneider}}, \citenamefont {{Le
  Brun}}, \citenamefont {{Beckmann}}, \citenamefont {{Peirani}}, \citenamefont
  {{Slyz}},\ and\ \citenamefont {{Pichon}}}]{2018MNRAS.480.3962C}%
  \BibitemOpen
  \bibfield  {author} {\bibinfo {author} {\bibfnamefont {N.~E.}\ \bibnamefont
  {{Chisari}}}, \bibinfo {author} {\bibfnamefont {M.~L.~A.}\ \bibnamefont
  {{Richardson}}}, \bibinfo {author} {\bibfnamefont {J.}~\bibnamefont
  {{Devriendt}}}, \bibinfo {author} {\bibfnamefont {Y.}~\bibnamefont
  {{Dubois}}}, \bibinfo {author} {\bibfnamefont {A.}~\bibnamefont
  {{Schneider}}}, \bibinfo {author} {\bibfnamefont {A.~M.~C.}\ \bibnamefont
  {{Le Brun}}}, \bibinfo {author} {\bibfnamefont {R.~S.}\ \bibnamefont
  {{Beckmann}}}, \bibinfo {author} {\bibfnamefont {S.}~\bibnamefont
  {{Peirani}}}, \bibinfo {author} {\bibfnamefont {A.}~\bibnamefont {{Slyz}}}, \
  and\ \bibinfo {author} {\bibfnamefont {C.}~\bibnamefont {{Pichon}}},\ }\href
  {\doibase 10.1093/mnras/sty2093} {\bibfield  {journal} {\bibinfo  {journal}
  {\mnras}\ }\textbf {\bibinfo {volume} {480}},\ \bibinfo {pages} {3962}
  (\bibinfo {year} {2018})},\ \Eprint {http://arxiv.org/abs/1801.08559}
  {arXiv:1801.08559} \BibitemShut {NoStop}%
\bibitem [{\citenamefont {Mehta}\ \emph {et~al.}(2011)\citenamefont {Mehta},
  \citenamefont {Seo}, \citenamefont {Eckel}, \citenamefont {Eisenstein},
  \citenamefont {Metchnik}, \citenamefont {Pinto},\ and\ \citenamefont
  {Xu}}]{0004-637X-734-2-94}%
  \BibitemOpen
  \bibfield  {author} {\bibinfo {author} {\bibfnamefont {K.~T.}\ \bibnamefont
  {Mehta}}, \bibinfo {author} {\bibfnamefont {H.-J.}\ \bibnamefont {Seo}},
  \bibinfo {author} {\bibfnamefont {J.}~\bibnamefont {Eckel}}, \bibinfo
  {author} {\bibfnamefont {D.~J.}\ \bibnamefont {Eisenstein}}, \bibinfo
  {author} {\bibfnamefont {M.}~\bibnamefont {Metchnik}}, \bibinfo {author}
  {\bibfnamefont {P.}~\bibnamefont {Pinto}}, \ and\ \bibinfo {author}
  {\bibfnamefont {X.}~\bibnamefont {Xu}},\ }\href
  {http://stacks.iop.org/0004-637X/734/i=2/a=94} {\bibfield  {journal}
  {\bibinfo  {journal} {The Astrophysical Journal}\ }\textbf {\bibinfo {volume}
  {734}},\ \bibinfo {pages} {94} (\bibinfo {year} {2011})}\BibitemShut
  {NoStop}%
\bibitem [{\citenamefont {Cunha}\ and\ \citenamefont
  {Evrard}(2010)}]{PhysRevD.81.083509}%
  \BibitemOpen
  \bibfield  {author} {\bibinfo {author} {\bibfnamefont {C.~E.}\ \bibnamefont
  {Cunha}}\ and\ \bibinfo {author} {\bibfnamefont {A.~E.}\ \bibnamefont
  {Evrard}},\ }\href {\doibase 10.1103/PhysRevD.81.083509} {\bibfield
  {journal} {\bibinfo  {journal} {Phys. Rev. D}\ }\textbf {\bibinfo {volume}
  {81}},\ \bibinfo {pages} {083509} (\bibinfo {year} {2010})}\BibitemShut
  {NoStop}%
\bibitem [{\citenamefont {{Joachimi}}\ \emph {et~al.}(2015)\citenamefont
  {{Joachimi}}, \citenamefont {{Cacciato}}, \citenamefont {{Kitching}},
  \citenamefont {{Leonard}}, \citenamefont {{Mandelbaum}}, \citenamefont
  {{Sch{\"a}fer}}, \citenamefont {{Sif{\'o}n}}, \citenamefont {{Hoekstra}},
  \citenamefont {{Kiessling}}, \citenamefont {{Kirk}},\ and\ \citenamefont
  {{Rassat}}}]{2015SSRv..193....1J}%
  \BibitemOpen
  \bibfield  {author} {\bibinfo {author} {\bibfnamefont {B.}~\bibnamefont
  {{Joachimi}}}, \bibinfo {author} {\bibfnamefont {M.}~\bibnamefont
  {{Cacciato}}}, \bibinfo {author} {\bibfnamefont {T.~D.}\ \bibnamefont
  {{Kitching}}}, \bibinfo {author} {\bibfnamefont {A.}~\bibnamefont
  {{Leonard}}}, \bibinfo {author} {\bibfnamefont {R.}~\bibnamefont
  {{Mandelbaum}}}, \bibinfo {author} {\bibfnamefont {B.~M.}\ \bibnamefont
  {{Sch{\"a}fer}}}, \bibinfo {author} {\bibfnamefont {C.}~\bibnamefont
  {{Sif{\'o}n}}}, \bibinfo {author} {\bibfnamefont {H.}~\bibnamefont
  {{Hoekstra}}}, \bibinfo {author} {\bibfnamefont {A.}~\bibnamefont
  {{Kiessling}}}, \bibinfo {author} {\bibfnamefont {D.}~\bibnamefont {{Kirk}}},
  \ and\ \bibinfo {author} {\bibfnamefont {A.}~\bibnamefont {{Rassat}}},\
  }\href {\doibase 10.1007/s11214-015-0177-4} {\bibfield  {journal} {\bibinfo
  {journal} {\ssr}\ }\textbf {\bibinfo {volume} {193}},\ \bibinfo {pages} {1}
  (\bibinfo {year} {2015})},\ \Eprint {http://arxiv.org/abs/1504.05456}
  {arXiv:1504.05456} \BibitemShut {NoStop}%
\bibitem [{\citenamefont {{Troxel}}\ and\ \citenamefont
  {{Ishak}}(2015)}]{2015PhR...558....1T}%
  \BibitemOpen
  \bibfield  {author} {\bibinfo {author} {\bibfnamefont {M.~A.}\ \bibnamefont
  {{Troxel}}}\ and\ \bibinfo {author} {\bibfnamefont {M.}~\bibnamefont
  {{Ishak}}},\ }\href {\doibase 10.1016/j.physrep.2014.11.001} {\bibfield
  {journal} {\bibinfo  {journal} {\physrep}\ }\textbf {\bibinfo {volume}
  {558}},\ \bibinfo {pages} {1} (\bibinfo {year} {2015})},\ \Eprint
  {http://arxiv.org/abs/1407.6990} {arXiv:1407.6990} \BibitemShut {NoStop}%
\bibitem [{\citenamefont {Morrison}\ and\ \citenamefont
  {Schneider}(2013)}]{Morrison:2013tqa}%
  \BibitemOpen
  \bibfield  {author} {\bibinfo {author} {\bibfnamefont {C.~B.}\ \bibnamefont
  {Morrison}}\ and\ \bibinfo {author} {\bibfnamefont {M.~D.}\ \bibnamefont
  {Schneider}},\ }\href {\doibase 10.1088/1475-7516/2013/11/009} {\bibfield
  {journal} {\bibinfo  {journal} {JCAP}\ }\textbf {\bibinfo {volume} {1311}},\
  \bibinfo {pages} {009} (\bibinfo {year} {2013})},\ \Eprint
  {http://arxiv.org/abs/1304.7789} {arXiv:1304.7789 [astro-ph.CO]} \BibitemShut
  {NoStop}%
%%CITATION = ARXIV:1304.7789;%%
\bibitem [{\citenamefont {Hall}\ and\ \citenamefont
  {Taylor}(2018)}]{Hall:2018umb}%
  \BibitemOpen
  \bibfield  {author} {\bibinfo {author} {\bibfnamefont {A.}~\bibnamefont
  {Hall}}\ and\ \bibinfo {author} {\bibfnamefont {A.}~\bibnamefont {Taylor}},\
  }\href@noop {} {\  (\bibinfo {year} {2018})},\ \Eprint
  {http://arxiv.org/abs/1807.06875} {arXiv:1807.06875 [astro-ph.CO]}
  \BibitemShut {NoStop}%
%%CITATION = ARXIV:1807.06875;%%
\bibitem [{\citenamefont {Dodelson}\ and\ \citenamefont
  {Schneider}(2013)}]{Dodelson:2013uaa}%
  \BibitemOpen
  \bibfield  {author} {\bibinfo {author} {\bibfnamefont {S.}~\bibnamefont
  {Dodelson}}\ and\ \bibinfo {author} {\bibfnamefont {M.~D.}\ \bibnamefont
  {Schneider}},\ }\href {\doibase 10.1103/PhysRevD.88.063537} {\bibfield
  {journal} {\bibinfo  {journal} {Phys. Rev.}\ }\textbf {\bibinfo {volume}
  {D88}},\ \bibinfo {pages} {063537} (\bibinfo {year} {2013})},\ \Eprint
  {http://arxiv.org/abs/1304.2593} {arXiv:1304.2593 [astro-ph.CO]} \BibitemShut
  {NoStop}%
%%CITATION = ARXIV:1304.2593;%%
\bibitem [{\citenamefont {O'Connell}\ and\ \citenamefont
  {Eisenstein}(2018)}]{OConnell:2018oqr}%
  \BibitemOpen
  \bibfield  {author} {\bibinfo {author} {\bibfnamefont {R.}~\bibnamefont
  {O'Connell}}\ and\ \bibinfo {author} {\bibfnamefont {D.~J.}\ \bibnamefont
  {Eisenstein}},\ }\href@noop {} {\  (\bibinfo {year} {2018})},\ \Eprint
  {http://arxiv.org/abs/1808.05978} {arXiv:1808.05978 [astro-ph.CO]}
  \BibitemShut {NoStop}%
%%CITATION = ARXIV:1808.05978;%%
\bibitem [{\citenamefont {Taylor}\ and\ \citenamefont
  {Joachimi}(2014)}]{Taylor:2014ota}%
  \BibitemOpen
  \bibfield  {author} {\bibinfo {author} {\bibfnamefont {A.}~\bibnamefont
  {Taylor}}\ and\ \bibinfo {author} {\bibfnamefont {B.}~\bibnamefont
  {Joachimi}},\ }\href {\doibase 10.1093/mnras/stu996} {\bibfield  {journal}
  {\bibinfo  {journal} {Mon. Not. Roy. Astron. Soc.}\ }\textbf {\bibinfo
  {volume} {442}},\ \bibinfo {pages} {2728} (\bibinfo {year} {2014})},\ \Eprint
  {http://arxiv.org/abs/1402.6983} {arXiv:1402.6983 [astro-ph.CO]} \BibitemShut
  {NoStop}%
%%CITATION = ARXIV:1402.6983;%%
\bibitem [{\citenamefont {Joachimi}\ and\ \citenamefont
  {Taylor}(2014)}]{Joachimi:2014hca}%
  \BibitemOpen
  \bibfield  {author} {\bibinfo {author} {\bibfnamefont {B.}~\bibnamefont
  {Joachimi}}\ and\ \bibinfo {author} {\bibfnamefont {A.}~\bibnamefont
  {Taylor}},\ }\bibfield  {booktitle} {\emph {\bibinfo {booktitle}
  {{Proceedings, IAU Symposium 306: Statistical Challenges in 21st Century
  Cosmology: Lisbon, Portugal, May 25-29, 2014}}},\ }\href {\doibase
  10.1017/S1743921314013428} {\bibfield  {journal} {\bibinfo  {journal} {IAU
  Symp.}\ }\textbf {\bibinfo {volume} {306}},\ \bibinfo {pages} {99} (\bibinfo
  {year} {2014})},\ \Eprint {http://arxiv.org/abs/1412.4914} {arXiv:1412.4914
  [astro-ph.IM]} \BibitemShut {NoStop}%
%%CITATION = ARXIV:1412.4914;%%
\bibitem [{\citenamefont {Paz}\ and\ \citenamefont
  {Sanchez}(2015)}]{Paz:2015kwa}%
  \BibitemOpen
  \bibfield  {author} {\bibinfo {author} {\bibfnamefont {D.~J.}\ \bibnamefont
  {Paz}}\ and\ \bibinfo {author} {\bibfnamefont {A.~G.}\ \bibnamefont
  {Sanchez}},\ }\href {\doibase 10.1093/mnras/stv2259} {\bibfield  {journal}
  {\bibinfo  {journal} {Mon. Not. Roy. Astron. Soc.}\ }\textbf {\bibinfo
  {volume} {454}},\ \bibinfo {pages} {4326} (\bibinfo {year} {2015})},\ \Eprint
  {http://arxiv.org/abs/1508.03162} {arXiv:1508.03162 [astro-ph.CO]}
  \BibitemShut {NoStop}%
%%CITATION = ARXIV:1508.03162;%%
\bibitem [{\citenamefont {Pearson}\ and\ \citenamefont
  {Samushia}(2016)}]{Pearson:2015gca}%
  \BibitemOpen
  \bibfield  {author} {\bibinfo {author} {\bibfnamefont {D.~W.}\ \bibnamefont
  {Pearson}}\ and\ \bibinfo {author} {\bibfnamefont {L.}~\bibnamefont
  {Samushia}},\ }\href {\doibase 10.1093/mnras/stw062} {\bibfield  {journal}
  {\bibinfo  {journal} {Mon. Not. Roy. Astron. Soc.}\ }\textbf {\bibinfo
  {volume} {457}},\ \bibinfo {pages} {993} (\bibinfo {year} {2016})},\ \Eprint
  {http://arxiv.org/abs/1509.00064} {arXiv:1509.00064 [astro-ph.CO]}
  \BibitemShut {NoStop}%
%%CITATION = ARXIV:1509.00064;%%
\bibitem [{\citenamefont {Petri}\ \emph {et~al.}(2016)\citenamefont {Petri},
  \citenamefont {Haiman},\ and\ \citenamefont {May}}]{Petri:2016wlu}%
  \BibitemOpen
  \bibfield  {author} {\bibinfo {author} {\bibfnamefont {A.}~\bibnamefont
  {Petri}}, \bibinfo {author} {\bibfnamefont {Z.}~\bibnamefont {Haiman}}, \
  and\ \bibinfo {author} {\bibfnamefont {M.}~\bibnamefont {May}},\ }\href
  {\doibase 10.1103/PhysRevD.93.063524} {\bibfield  {journal} {\bibinfo
  {journal} {Phys. Rev.}\ }\textbf {\bibinfo {volume} {D93}},\ \bibinfo {pages}
  {063524} (\bibinfo {year} {2016})},\ \Eprint
  {http://arxiv.org/abs/1601.06792} {arXiv:1601.06792 [astro-ph.CO]}
  \BibitemShut {NoStop}%
%%CITATION = ARXIV:1601.06792;%%
\bibitem [{\citenamefont {Heavens}\ \emph {et~al.}(2017)\citenamefont
  {Heavens}, \citenamefont {Sellentin}, \citenamefont {de~Mijolla},\ and\
  \citenamefont {Vianello}}]{Heavens:2017efz}%
  \BibitemOpen
  \bibfield  {author} {\bibinfo {author} {\bibfnamefont {A.}~\bibnamefont
  {Heavens}}, \bibinfo {author} {\bibfnamefont {E.}~\bibnamefont {Sellentin}},
  \bibinfo {author} {\bibfnamefont {D.}~\bibnamefont {de~Mijolla}}, \ and\
  \bibinfo {author} {\bibfnamefont {A.}~\bibnamefont {Vianello}},\ }\href
  {\doibase 10.1093/mnras/stx2326} {\bibfield  {journal} {\bibinfo  {journal}
  {Mon. Not. Roy. Astron. Soc.}\ }\textbf {\bibinfo {volume} {472}},\ \bibinfo
  {pages} {4244} (\bibinfo {year} {2017})},\ \Eprint
  {http://arxiv.org/abs/1707.06529} {arXiv:1707.06529 [astro-ph.CO]}
  \BibitemShut {NoStop}%
%%CITATION = ARXIV:1707.06529;%%
\bibitem [{\citenamefont {Sellentin}\ and\ \citenamefont
  {Heavens}(2016)}]{Sellentin:2015waz}%
  \BibitemOpen
  \bibfield  {author} {\bibinfo {author} {\bibfnamefont {E.}~\bibnamefont
  {Sellentin}}\ and\ \bibinfo {author} {\bibfnamefont {A.~F.}\ \bibnamefont
  {Heavens}},\ }\href {\doibase 10.1093/mnrasl/slv190} {\bibfield  {journal}
  {\bibinfo  {journal} {Mon. Not. Roy. Astron. Soc.}\ }\textbf {\bibinfo
  {volume} {456}},\ \bibinfo {pages} {L132} (\bibinfo {year} {2016})},\ \Eprint
  {http://arxiv.org/abs/1511.05969} {arXiv:1511.05969 [astro-ph.CO]}
  \BibitemShut {NoStop}%
%%CITATION = ARXIV:1511.05969;%%
\bibitem [{\citenamefont {Sellentin}\ and\ \citenamefont
  {Heavens}(2017)}]{Sellentin:2016psv}%
  \BibitemOpen
  \bibfield  {author} {\bibinfo {author} {\bibfnamefont {E.}~\bibnamefont
  {Sellentin}}\ and\ \bibinfo {author} {\bibfnamefont {A.~F.}\ \bibnamefont
  {Heavens}},\ }\href {\doibase 10.1093/mnras/stw2697} {\bibfield  {journal}
  {\bibinfo  {journal} {Mon. Not. Roy. Astron. Soc.}\ }\textbf {\bibinfo
  {volume} {464}},\ \bibinfo {pages} {4658} (\bibinfo {year} {2017})},\ \Eprint
  {http://arxiv.org/abs/1609.00504} {arXiv:1609.00504 [astro-ph.CO]}
  \BibitemShut {NoStop}%
%%CITATION = ARXIV:1609.00504;%%
\bibitem [{\citenamefont {Mohammed}\ \emph {et~al.}(2017)\citenamefont
  {Mohammed}, \citenamefont {Seljak},\ and\ \citenamefont
  {Vlah}}]{Mohammed:2016sre}%
  \BibitemOpen
  \bibfield  {author} {\bibinfo {author} {\bibfnamefont {I.}~\bibnamefont
  {Mohammed}}, \bibinfo {author} {\bibfnamefont {U.}~\bibnamefont {Seljak}}, \
  and\ \bibinfo {author} {\bibfnamefont {Z.}~\bibnamefont {Vlah}},\ }\href
  {\doibase 10.1093/mnras/stw3196} {\bibfield  {journal} {\bibinfo  {journal}
  {Mon. Not. Roy. Astron. Soc.}\ }\textbf {\bibinfo {volume} {466}},\ \bibinfo
  {pages} {780} (\bibinfo {year} {2017})},\ \Eprint
  {http://arxiv.org/abs/1607.00043} {arXiv:1607.00043 [astro-ph.CO]}
  \BibitemShut {NoStop}%
%%CITATION = ARXIV:1607.00043;%%
\bibitem [{\citenamefont {Barreira}\ and\ \citenamefont
  {Schmidt}(2017{\natexlab{a}})}]{Barreira:2017sqa}%
  \BibitemOpen
  \bibfield  {author} {\bibinfo {author} {\bibfnamefont {A.}~\bibnamefont
  {Barreira}}\ and\ \bibinfo {author} {\bibfnamefont {F.}~\bibnamefont
  {Schmidt}},\ }\href {\doibase 10.1088/1475-7516/2017/06/053} {\bibfield
  {journal} {\bibinfo  {journal} {JCAP}\ }\textbf {\bibinfo {volume} {1706}},\
  \bibinfo {pages} {053} (\bibinfo {year} {2017}{\natexlab{a}})},\ \Eprint
  {http://arxiv.org/abs/1703.09212} {arXiv:1703.09212 [astro-ph.CO]}
  \BibitemShut {NoStop}%
%%CITATION = ARXIV:1703.09212;%%
\bibitem [{\citenamefont {Barreira}\ and\ \citenamefont
  {Schmidt}(2017{\natexlab{b}})}]{Barreira:2017kxd}%
  \BibitemOpen
  \bibfield  {author} {\bibinfo {author} {\bibfnamefont {A.}~\bibnamefont
  {Barreira}}\ and\ \bibinfo {author} {\bibfnamefont {F.}~\bibnamefont
  {Schmidt}},\ }\href {\doibase 10.1088/1475-7516/2017/11/051} {\bibfield
  {journal} {\bibinfo  {journal} {JCAP}\ }\textbf {\bibinfo {volume} {1711}},\
  \bibinfo {pages} {051} (\bibinfo {year} {2017}{\natexlab{b}})},\ \Eprint
  {http://arxiv.org/abs/1705.01092} {arXiv:1705.01092 [astro-ph.CO]}
  \BibitemShut {NoStop}%
%%CITATION = ARXIV:1705.01092;%%
\bibitem [{\citenamefont {Barreira}\ \emph {et~al.}(2018)\citenamefont
  {Barreira}, \citenamefont {Krause},\ and\ \citenamefont
  {Schmidt}}]{Barreira:2018jgd}%
  \BibitemOpen
  \bibfield  {author} {\bibinfo {author} {\bibfnamefont {A.}~\bibnamefont
  {Barreira}}, \bibinfo {author} {\bibfnamefont {E.}~\bibnamefont {Krause}}, \
  and\ \bibinfo {author} {\bibfnamefont {F.}~\bibnamefont {Schmidt}},\ }\href
  {\doibase 10.1088/1475-7516/2018/10/053} {\bibfield  {journal} {\bibinfo
  {journal} {JCAP}\ }\textbf {\bibinfo {volume} {1810}},\ \bibinfo {pages}
  {053} (\bibinfo {year} {2018})},\ \Eprint {http://arxiv.org/abs/1807.04266}
  {arXiv:1807.04266 [astro-ph.CO]} \BibitemShut {NoStop}%
%%CITATION = ARXIV:1807.04266;%%
\bibitem [{\citenamefont {Li}\ \emph {et~al.}(2016)\citenamefont {Li},
  \citenamefont {Hu},\ and\ \citenamefont {Takada}}]{Li:2015jsz}%
  \BibitemOpen
  \bibfield  {author} {\bibinfo {author} {\bibfnamefont {Y.}~\bibnamefont
  {Li}}, \bibinfo {author} {\bibfnamefont {W.}~\bibnamefont {Hu}}, \ and\
  \bibinfo {author} {\bibfnamefont {M.}~\bibnamefont {Takada}},\ }\href
  {\doibase 10.1103/PhysRevD.93.063507} {\bibfield  {journal} {\bibinfo
  {journal} {Phys. Rev.}\ }\textbf {\bibinfo {volume} {D93}},\ \bibinfo {pages}
  {063507} (\bibinfo {year} {2016})},\ \Eprint
  {http://arxiv.org/abs/1511.01454} {arXiv:1511.01454 [astro-ph.CO]}
  \BibitemShut {NoStop}%
%%CITATION = ARXIV:1511.01454;%%
\bibitem [{\citenamefont {Wagner}\ \emph {et~al.}(2015)\citenamefont {Wagner},
  \citenamefont {Schmidt}, \citenamefont {Chiang},\ and\ \citenamefont
  {Komatsu}}]{Wagner:2015gva}%
  \BibitemOpen
  \bibfield  {author} {\bibinfo {author} {\bibfnamefont {C.}~\bibnamefont
  {Wagner}}, \bibinfo {author} {\bibfnamefont {F.}~\bibnamefont {Schmidt}},
  \bibinfo {author} {\bibfnamefont {C.-T.}\ \bibnamefont {Chiang}}, \ and\
  \bibinfo {author} {\bibfnamefont {E.}~\bibnamefont {Komatsu}},\ }\href
  {\doibase 10.1088/1475-7516/2015/08/042} {\bibfield  {journal} {\bibinfo
  {journal} {JCAP}\ }\textbf {\bibinfo {volume} {1508}},\ \bibinfo {pages}
  {042} (\bibinfo {year} {2015})},\ \Eprint {http://arxiv.org/abs/1503.03487}
  {arXiv:1503.03487 [astro-ph.CO]} \BibitemShut {NoStop}%
%%CITATION = ARXIV:1503.03487;%%
\bibitem [{\citenamefont {{Carron}}(2013)}]{2013A&A...551A..88C}%
  \BibitemOpen
  \bibfield  {author} {\bibinfo {author} {\bibfnamefont {J.}~\bibnamefont
  {{Carron}}},\ }\href {\doibase 10.1051/0004-6361/201220538} {\bibfield
  {journal} {\bibinfo  {journal} {\aap}\ }\textbf {\bibinfo {volume} {551}},\
  \bibinfo {eid} {A88} (\bibinfo {year} {2013})},\ \Eprint
  {http://arxiv.org/abs/1204.4724} {arXiv:1204.4724 [astro-ph.CO]} \BibitemShut
  {NoStop}%
\bibitem [{\citenamefont {Eifler}\ \emph {et~al.}(2009)\citenamefont {Eifler},
  \citenamefont {Schneider},\ and\ \citenamefont {Hartlap}}]{Eifler:2008gx}%
  \BibitemOpen
  \bibfield  {author} {\bibinfo {author} {\bibfnamefont {T.}~\bibnamefont
  {Eifler}}, \bibinfo {author} {\bibfnamefont {P.}~\bibnamefont {Schneider}}, \
  and\ \bibinfo {author} {\bibfnamefont {J.}~\bibnamefont {Hartlap}},\ }\href
  {\doibase 10.1051/0004-6361/200811276} {\bibfield  {journal} {\bibinfo
  {journal} {Astron. Astrophys.}\ }\textbf {\bibinfo {volume} {502}},\ \bibinfo
  {pages} {721} (\bibinfo {year} {2009})},\ \Eprint
  {http://arxiv.org/abs/0810.4254} {arXiv:0810.4254 [astro-ph]} \BibitemShut
  {NoStop}%
%%CITATION = ARXIV:0810.4254;%%
\bibitem [{\citenamefont {{Amara}}\ and\ \citenamefont
  {{R{\'e}fr{\'e}gier}}(2008)}]{2008MNRAS.391..228A}%
  \BibitemOpen
  \bibfield  {author} {\bibinfo {author} {\bibfnamefont {A.}~\bibnamefont
  {{Amara}}}\ and\ \bibinfo {author} {\bibfnamefont {A.}~\bibnamefont
  {{R{\'e}fr{\'e}gier}}},\ }\href {\doibase 10.1111/j.1365-2966.2008.13880.x}
  {\bibfield  {journal} {\bibinfo  {journal} {\mnras}\ }\textbf {\bibinfo
  {volume} {391}},\ \bibinfo {pages} {228} (\bibinfo {year} {2008})},\ \Eprint
  {http://arxiv.org/abs/0710.5171} {arXiv:0710.5171} \BibitemShut {NoStop}%
\bibitem [{\citenamefont {{Taburet}}\ \emph {et~al.}(2009)\citenamefont
  {{Taburet}}, \citenamefont {{Aghanim}}, \citenamefont {{Douspis}},\ and\
  \citenamefont {{Langer}}}]{2009MNRAS.392.1153T}%
  \BibitemOpen
  \bibfield  {author} {\bibinfo {author} {\bibfnamefont {N.}~\bibnamefont
  {{Taburet}}}, \bibinfo {author} {\bibfnamefont {N.}~\bibnamefont
  {{Aghanim}}}, \bibinfo {author} {\bibfnamefont {M.}~\bibnamefont
  {{Douspis}}}, \ and\ \bibinfo {author} {\bibfnamefont {M.}~\bibnamefont
  {{Langer}}},\ }\href {\doibase 10.1111/j.1365-2966.2008.14105.x} {\bibfield
  {journal} {\bibinfo  {journal} {\mnras}\ }\textbf {\bibinfo {volume} {392}},\
  \bibinfo {pages} {1153} (\bibinfo {year} {2009})},\ \Eprint
  {http://arxiv.org/abs/0809.1364} {arXiv:0809.1364} \BibitemShut {NoStop}%
\bibitem [{\citenamefont {{Reischke}}\ \emph {et~al.}(2017)\citenamefont
  {{Reischke}}, \citenamefont {{Kiessling}},\ and\ \citenamefont
  {{Sch{\"a}fer}}}]{2017MNRAS.465.4016R}%
  \BibitemOpen
  \bibfield  {author} {\bibinfo {author} {\bibfnamefont {R.}~\bibnamefont
  {{Reischke}}}, \bibinfo {author} {\bibfnamefont {A.}~\bibnamefont
  {{Kiessling}}}, \ and\ \bibinfo {author} {\bibfnamefont {B.~M.}\ \bibnamefont
  {{Sch{\"a}fer}}},\ }\href {\doibase 10.1093/mnras/stw2976} {\bibfield
  {journal} {\bibinfo  {journal} {\mnras}\ }\textbf {\bibinfo {volume} {465}},\
  \bibinfo {pages} {4016} (\bibinfo {year} {2017})},\ \Eprint
  {http://arxiv.org/abs/1607.03136} {arXiv:1607.03136 [astro-ph.CO]}
  \BibitemShut {NoStop}%
\bibitem [{\citenamefont {Hamimeche}\ and\ \citenamefont
  {Lewis}(2008)}]{Hamimeche:2008ai}%
  \BibitemOpen
  \bibfield  {author} {\bibinfo {author} {\bibfnamefont {S.}~\bibnamefont
  {Hamimeche}}\ and\ \bibinfo {author} {\bibfnamefont {A.}~\bibnamefont
  {Lewis}},\ }\href {\doibase 10.1103/PhysRevD.77.103013} {\bibfield  {journal}
  {\bibinfo  {journal} {Phys. Rev.}\ }\textbf {\bibinfo {volume} {D77}},\
  \bibinfo {pages} {103013} (\bibinfo {year} {2008})},\ \Eprint
  {http://arxiv.org/abs/0801.0554} {arXiv:0801.0554 [astro-ph]} \BibitemShut
  {NoStop}%
%%CITATION = ARXIV:0801.0554;%%
\bibitem [{\citenamefont {{Tegmark}}\ \emph {et~al.}(1997)\citenamefont
  {{Tegmark}}, \citenamefont {{Taylor}},\ and\ \citenamefont
  {{Heavens}}}]{1997ApJ...480...22T}%
  \BibitemOpen
  \bibfield  {author} {\bibinfo {author} {\bibfnamefont {M.}~\bibnamefont
  {{Tegmark}}}, \bibinfo {author} {\bibfnamefont {A.~N.}\ \bibnamefont
  {{Taylor}}}, \ and\ \bibinfo {author} {\bibfnamefont {A.~F.}\ \bibnamefont
  {{Heavens}}},\ }\href {\doibase 10.1086/303939} {\bibfield  {journal}
  {\bibinfo  {journal} {\apj}\ }\textbf {\bibinfo {volume} {480}},\ \bibinfo
  {pages} {22} (\bibinfo {year} {1997})},\ \Eprint
  {http://arxiv.org/abs/astro-ph/9603021} {astro-ph/9603021} \BibitemShut
  {NoStop}%
\bibitem [{\citenamefont {{Simon}}(2007)}]{2007A&A...473..711S}%
  \BibitemOpen
  \bibfield  {author} {\bibinfo {author} {\bibfnamefont {P.}~\bibnamefont
  {{Simon}}},\ }\href {\doibase 10.1051/0004-6361:20066352} {\bibfield
  {journal} {\bibinfo  {journal} {\aap}\ }\textbf {\bibinfo {volume} {473}},\
  \bibinfo {pages} {711} (\bibinfo {year} {2007})},\ \Eprint
  {http://arxiv.org/abs/astro-ph/0609165} {astro-ph/0609165} \BibitemShut
  {NoStop}%
\bibitem [{\citenamefont {{Bartelmann}}\ and\ \citenamefont
  {{Schneider}}(2001)}]{2001PhR...340..291B}%
  \BibitemOpen
  \bibfield  {author} {\bibinfo {author} {\bibfnamefont {M.}~\bibnamefont
  {{Bartelmann}}}\ and\ \bibinfo {author} {\bibfnamefont {P.}~\bibnamefont
  {{Schneider}}},\ }\href {\doibase 10.1016/S0370-1573(00)00082-X} {\bibfield
  {journal} {\bibinfo  {journal} {\physrep}\ }\textbf {\bibinfo {volume}
  {340}},\ \bibinfo {pages} {291} (\bibinfo {year} {2001})},\ \Eprint
  {http://arxiv.org/abs/astro-ph/9912508} {astro-ph/9912508} \BibitemShut
  {NoStop}%
\bibitem [{\citenamefont {Takada}\ and\ \citenamefont
  {Hu}(2013)}]{Takada:2013bfn}%
  \BibitemOpen
  \bibfield  {author} {\bibinfo {author} {\bibfnamefont {M.}~\bibnamefont
  {Takada}}\ and\ \bibinfo {author} {\bibfnamefont {W.}~\bibnamefont {Hu}},\
  }\href {\doibase 10.1103/PhysRevD.87.123504} {\bibfield  {journal} {\bibinfo
  {journal} {Phys. Rev.}\ }\textbf {\bibinfo {volume} {D87}},\ \bibinfo {pages}
  {123504} (\bibinfo {year} {2013})},\ \Eprint {http://arxiv.org/abs/1302.6994}
  {arXiv:1302.6994 [astro-ph.CO]} \BibitemShut {NoStop}%
%%CITATION = ARXIV:1302.6994;%%
\bibitem [{\citenamefont {{Knox}}(1995)}]{1995PhRvD..52.4307K}%
  \BibitemOpen
  \bibfield  {author} {\bibinfo {author} {\bibfnamefont {L.}~\bibnamefont
  {{Knox}}},\ }\href {\doibase 10.1103/PhysRevD.52.4307} {\bibfield  {journal}
  {\bibinfo  {journal} {\prd}\ }\textbf {\bibinfo {volume} {52}},\ \bibinfo
  {pages} {4307} (\bibinfo {year} {1995})},\ \Eprint
  {http://arxiv.org/abs/astro-ph/9504054} {astro-ph/9504054} \BibitemShut
  {NoStop}%
\bibitem [{\citenamefont {{Crocce}}\ \emph {et~al.}(2011)\citenamefont
  {{Crocce}}, \citenamefont {{Cabr{\'e}}},\ and\ \citenamefont
  {{Gazta{\~n}aga}}}]{2011MNRAS.414..329C}%
  \BibitemOpen
  \bibfield  {author} {\bibinfo {author} {\bibfnamefont {M.}~\bibnamefont
  {{Crocce}}}, \bibinfo {author} {\bibfnamefont {A.}~\bibnamefont
  {{Cabr{\'e}}}}, \ and\ \bibinfo {author} {\bibfnamefont {E.}~\bibnamefont
  {{Gazta{\~n}aga}}},\ }\href {\doibase 10.1111/j.1365-2966.2011.18393.x}
  {\bibfield  {journal} {\bibinfo  {journal} {\mnras}\ }\textbf {\bibinfo
  {volume} {414}},\ \bibinfo {pages} {329} (\bibinfo {year} {2011})},\ \Eprint
  {http://arxiv.org/abs/1004.4640} {arXiv:1004.4640} \BibitemShut {NoStop}%
\bibitem [{\citenamefont {{The LSST Dark Energy Science Collaboration}}\ \emph
  {et~al.}(2018)\citenamefont {{The LSST Dark Energy Science Collaboration}}
  \emph {et~al.}}]{2018arXiv180901669T}%
  \BibitemOpen
  \bibfield  {author} {\bibinfo {author} {\bibnamefont {{The LSST Dark Energy
  Science Collaboration}}} \emph {et~al.},\ }\href@noop {} {\bibfield
  {journal} {\bibinfo  {journal} {ArXiv e-prints}\ } (\bibinfo {year}
  {2018})},\ \Eprint {http://arxiv.org/abs/1809.01669} {arXiv:1809.01669}
  \BibitemShut {NoStop}%
\bibitem [{\citenamefont {{Weinberg}}\ \emph {et~al.}(2004)\citenamefont
  {{Weinberg}}, \citenamefont {{Dav{\'e}}}, \citenamefont {{Katz}},\ and\
  \citenamefont {{Hernquist}}}]{2004ApJ...601....1W}%
  \BibitemOpen
  \bibfield  {author} {\bibinfo {author} {\bibfnamefont {D.~H.}\ \bibnamefont
  {{Weinberg}}}, \bibinfo {author} {\bibfnamefont {R.}~\bibnamefont
  {{Dav{\'e}}}}, \bibinfo {author} {\bibfnamefont {N.}~\bibnamefont {{Katz}}},
  \ and\ \bibinfo {author} {\bibfnamefont {L.}~\bibnamefont {{Hernquist}}},\
  }\href {\doibase 10.1086/380481} {\bibfield  {journal} {\bibinfo  {journal}
  {\apj}\ }\textbf {\bibinfo {volume} {601}},\ \bibinfo {pages} {1} (\bibinfo
  {year} {2004})},\ \Eprint {http://arxiv.org/abs/astro-ph/0212356}
  {astro-ph/0212356} \BibitemShut {NoStop}%
\bibitem [{\citenamefont {{The LSST Dark Energy Science Collaboration}}\ \emph
  {et~al.}()\citenamefont {{The LSST Dark Energy Science Collaboration}} \emph
  {et~al.}}]{ccl}%
  \BibitemOpen
  \bibfield  {author} {\bibinfo {author} {\bibnamefont {{The LSST Dark Energy
  Science Collaboration}}} \emph {et~al.},\ }\href@noop {} {\bibinfo  {journal}
  {In preparation.}\ }\BibitemShut {NoStop}%
\bibitem [{\citenamefont {{Erben}}\ \emph {et~al.}(2013)\citenamefont {{Erben}}
  \emph {et~al.}}]{2013MNRAS.433.2545E}%
  \BibitemOpen
\bibfield  {journal} {  }\bibfield  {author} {\bibinfo {author} {\bibfnamefont
  {T.}~\bibnamefont {{Erben}}} \emph {et~al.},\ }\href {\doibase
  10.1093/mnras/stt928} {\bibfield  {journal} {\bibinfo  {journal} {\mnras}\
  }\textbf {\bibinfo {volume} {433}},\ \bibinfo {pages} {2545} (\bibinfo {year}
  {2013})},\ \Eprint {http://arxiv.org/abs/1210.8156} {arXiv:1210.8156}
  \BibitemShut {NoStop}%
\bibitem [{\citenamefont {Krause}\ and\ \citenamefont
  {Eifler}(2017)}]{Krause:2016jvl}%
  \BibitemOpen
  \bibfield  {author} {\bibinfo {author} {\bibfnamefont {E.}~\bibnamefont
  {Krause}}\ and\ \bibinfo {author} {\bibfnamefont {T.}~\bibnamefont
  {Eifler}},\ }\href {\doibase 10.1093/mnras/stx1261} {\bibfield  {journal}
  {\bibinfo  {journal} {Mon. Not. Roy. Astron. Soc.}\ }\textbf {\bibinfo
  {volume} {470}},\ \bibinfo {pages} {2100} (\bibinfo {year} {2017})},\ \Eprint
  {http://arxiv.org/abs/1601.05779} {arXiv:1601.05779 [astro-ph.CO]}
  \BibitemShut {NoStop}%
%%CITATION = ARXIV:1601.05779;%%
\bibitem [{\citenamefont {Li}\ \emph {et~al.}(2014)\citenamefont {Li},
  \citenamefont {Hu},\ and\ \citenamefont {Takada}}]{Li:2014sga}%
  \BibitemOpen
  \bibfield  {author} {\bibinfo {author} {\bibfnamefont {Y.}~\bibnamefont
  {Li}}, \bibinfo {author} {\bibfnamefont {W.}~\bibnamefont {Hu}}, \ and\
  \bibinfo {author} {\bibfnamefont {M.}~\bibnamefont {Takada}},\ }\href
  {\doibase 10.1103/PhysRevD.89.083519} {\bibfield  {journal} {\bibinfo
  {journal} {Phys. Rev.}\ }\textbf {\bibinfo {volume} {D89}},\ \bibinfo {pages}
  {083519} (\bibinfo {year} {2014})},\ \Eprint {http://arxiv.org/abs/1401.0385}
  {arXiv:1401.0385 [astro-ph.CO]} \BibitemShut {NoStop}%
%%CITATION = ARXIV:1401.0385;%%
\bibitem [{\citenamefont {Hamilton}\ \emph {et~al.}(2006)\citenamefont
  {Hamilton}, \citenamefont {Rimes},\ and\ \citenamefont
  {Scoccimarro}}]{Hamilton:2005dx}%
  \BibitemOpen
  \bibfield  {author} {\bibinfo {author} {\bibfnamefont {A.~J.~S.}\
  \bibnamefont {Hamilton}}, \bibinfo {author} {\bibfnamefont {C.~D.}\
  \bibnamefont {Rimes}}, \ and\ \bibinfo {author} {\bibfnamefont
  {R.}~\bibnamefont {Scoccimarro}},\ }\href {\doibase
  10.1111/j.1365-2966.2006.10709.x} {\bibfield  {journal} {\bibinfo  {journal}
  {Mon. Not. Roy. Astron. Soc.}\ }\textbf {\bibinfo {volume} {371}},\ \bibinfo
  {pages} {1188} (\bibinfo {year} {2006})},\ \Eprint
  {http://arxiv.org/abs/astro-ph/0511416} {arXiv:astro-ph/0511416 [astro-ph]}
  \BibitemShut {NoStop}%
%%CITATION = ASTRO-PH/0511416;%%
\bibitem [{\citenamefont {{Tinker}}\ \emph {et~al.}(2008)\citenamefont
  {{Tinker}}, \citenamefont {{Kravtsov}}, \citenamefont {{Klypin}},
  \citenamefont {{Abazajian}}, \citenamefont {{Warren}}, \citenamefont
  {{Yepes}}, \citenamefont {{Gottl{\"o}ber}},\ and\ \citenamefont
  {{Holz}}}]{2008ApJ...688..709T}%
  \BibitemOpen
  \bibfield  {author} {\bibinfo {author} {\bibfnamefont {J.}~\bibnamefont
  {{Tinker}}}, \bibinfo {author} {\bibfnamefont {A.~V.}\ \bibnamefont
  {{Kravtsov}}}, \bibinfo {author} {\bibfnamefont {A.}~\bibnamefont
  {{Klypin}}}, \bibinfo {author} {\bibfnamefont {K.}~\bibnamefont
  {{Abazajian}}}, \bibinfo {author} {\bibfnamefont {M.}~\bibnamefont
  {{Warren}}}, \bibinfo {author} {\bibfnamefont {G.}~\bibnamefont {{Yepes}}},
  \bibinfo {author} {\bibfnamefont {S.}~\bibnamefont {{Gottl{\"o}ber}}}, \ and\
  \bibinfo {author} {\bibfnamefont {D.~E.}\ \bibnamefont {{Holz}}},\ }\href
  {\doibase 10.1086/591439} {\bibfield  {journal} {\bibinfo  {journal} {\apj}\
  }\textbf {\bibinfo {volume} {688}},\ \bibinfo {pages} {709} (\bibinfo {year}
  {2008})},\ \Eprint {http://arxiv.org/abs/0803.2706} {arXiv:0803.2706}
  \BibitemShut {NoStop}%
\end{thebibliography}%

  \appendix
  \section{Computing the SSC using the Halo model}\label{app:SSC}
    This appendix presents a complete description of the steps taken to estimate the super-sample contribution to the large-scale structure covariance matrix. 
    
    We follow the description in \cite{Krause:2016jvl}, and write the SSC in terms of the window functions of the tracers involved
    \begin{align}\label{eq:ssc_full}
      &\left(\covar_{\rm SSC}\right)^{ij,\ell_1}_{mn,\ell_2}=\int d\chi\frac{q^i(\chi)q^j(\chi)q^m(\chi)q^n(\chi)}{\chi^4}\times\\\nonumber
      &\hspace{12pt}\times {\cal R}(k_{\ell_1},z){\cal R}(k_{\ell_2},z)\,P(k_{\ell_1},z)P(k_{\ell_2},z)
      \,\sigma_b(f_{\rm sky},z), 
    \end{align}
    where the window functions $q^p$ are given by Eqs. \ref{eq:q_delta} and \ref{eq:q_lensing} for galaxy clustering and weak lensing respectively, $k_\ell\equiv(\ell+1/2)/\chi$, $z$ is the redshift at the comoving radial distance $\chi$ and $P(k,z)$ is the matter power spectrum. ${\cal R}(k,z)$ is the response of the matter power spectrum to a long-wavelength density perturbation $\delta_b$
    \begin{figure}
      \centering
      \includegraphics[width=0.47\textwidth]{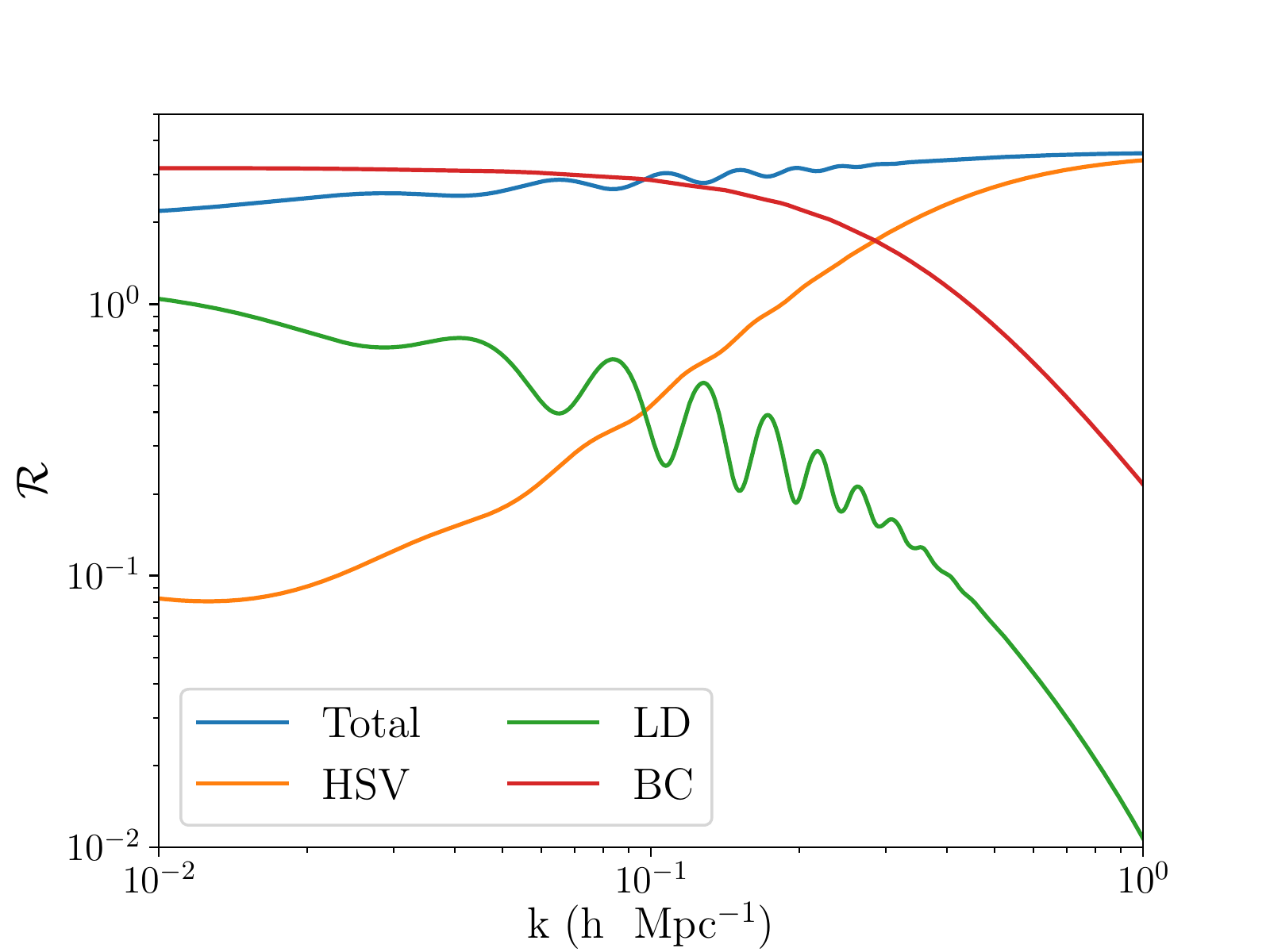}
      \caption{Different contributions to the power spectrum response at $z=0$ computed using the halo model (see Equation \ref{eq:response_halomod})}
      \label{fig:Response}
    \end{figure}
    \begin{figure}
      \begin{center}
        \includegraphics[width=0.47\textwidth]{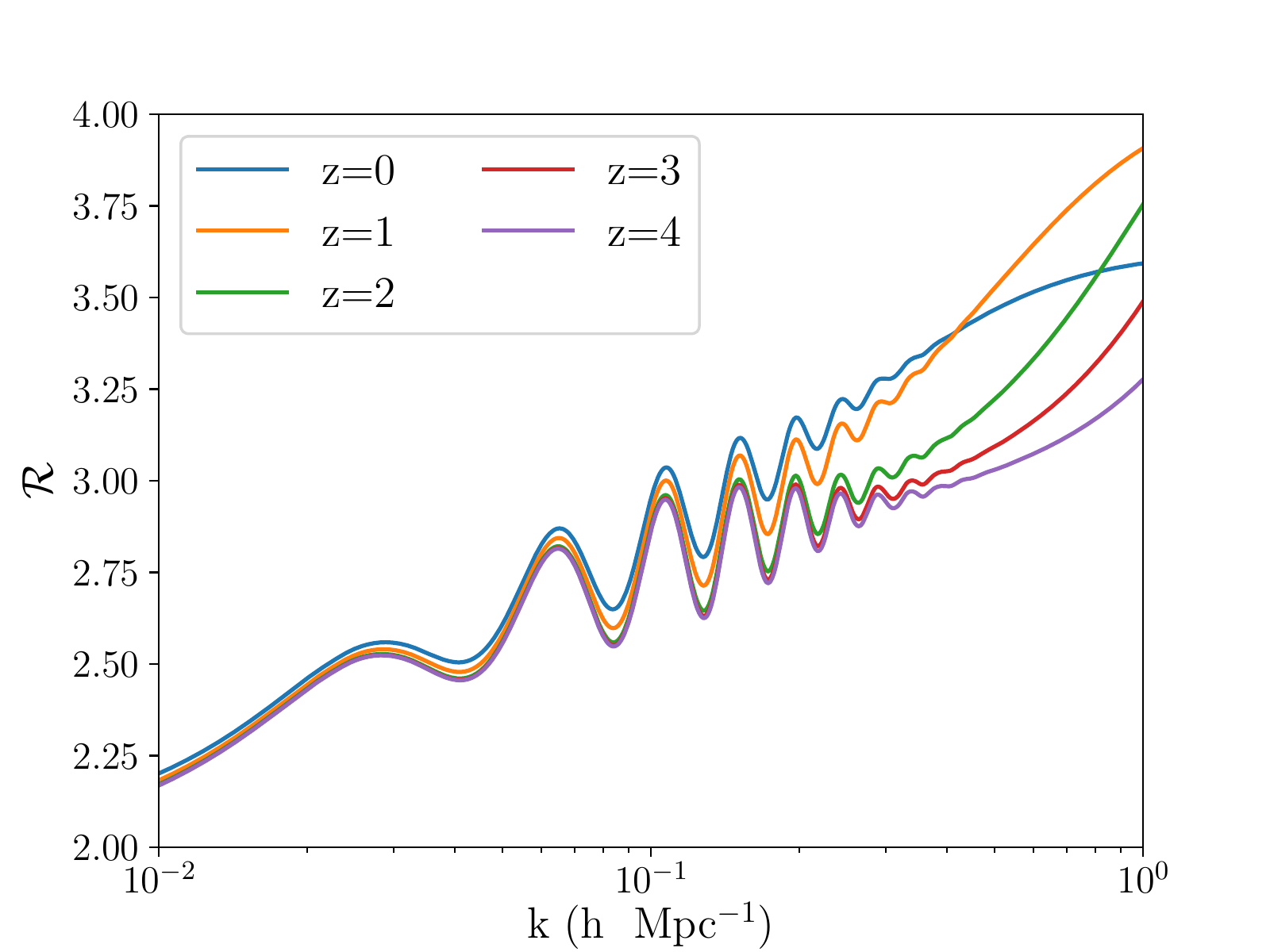}
        \caption{Power spectrum response as a function of redshift computing using the halo model.}
        \label{fig:Response_z}
      \end{center}
    \end{figure}
    \begin{equation}
      {\cal R}(k,z)\equiv\frac{d\log P(k,z)}{d\delta_b},
    \end{equation}
    and $\sigma_b$ is the projected variance of the density field within the footprint of your survey. For simplicity we assume a connected disc with an area corresponding to the sky fraction covered, in which case $\sigma_b$ can be estimated as
    \begin{equation}
     \sigma_b(f_{\rm sky},z)=\int dk_\perp^2 P_L(k_\perp,z)\left|\frac{2J_1(k_\perp R_s)}{k_\perp R_s}\right|^2,
    \end{equation}
    where $P_L$ is the linear matter power spectrum, $J_1$ is the order-1 cylindrical Bessel function and
    \begin{equation}
      R_s\equiv \chi(z)\,\theta_s,\hspace{12pt}\theta_s\equiv{\rm arccos}(1-2f_{\rm sky}).
    \end{equation}
    The expressions above reproduce the results of previous work by \cite{Krause:2016jvl}, in the simplified scenario of linear galaxy bias. We refer the reader to these works for further details.
    
    A standard approach to compute the power spectrum response ${\cal R}$ is to make use of so-called ``Separate-Universe'' simulations \cite{Li:2014sga, Hamilton:2005dx} in which the effects of a long-wavelength mode are modelled as modifying the effective background cosmology. Since our results will not depend strongly on the details of this calculation, we proceed as in \cite{Krause:2016jvl} and estimate ${\cal R}$ using a halo model approach. 
    
    \begin{figure}
      \begin{center}
        \includegraphics[width=0.49\textwidth]{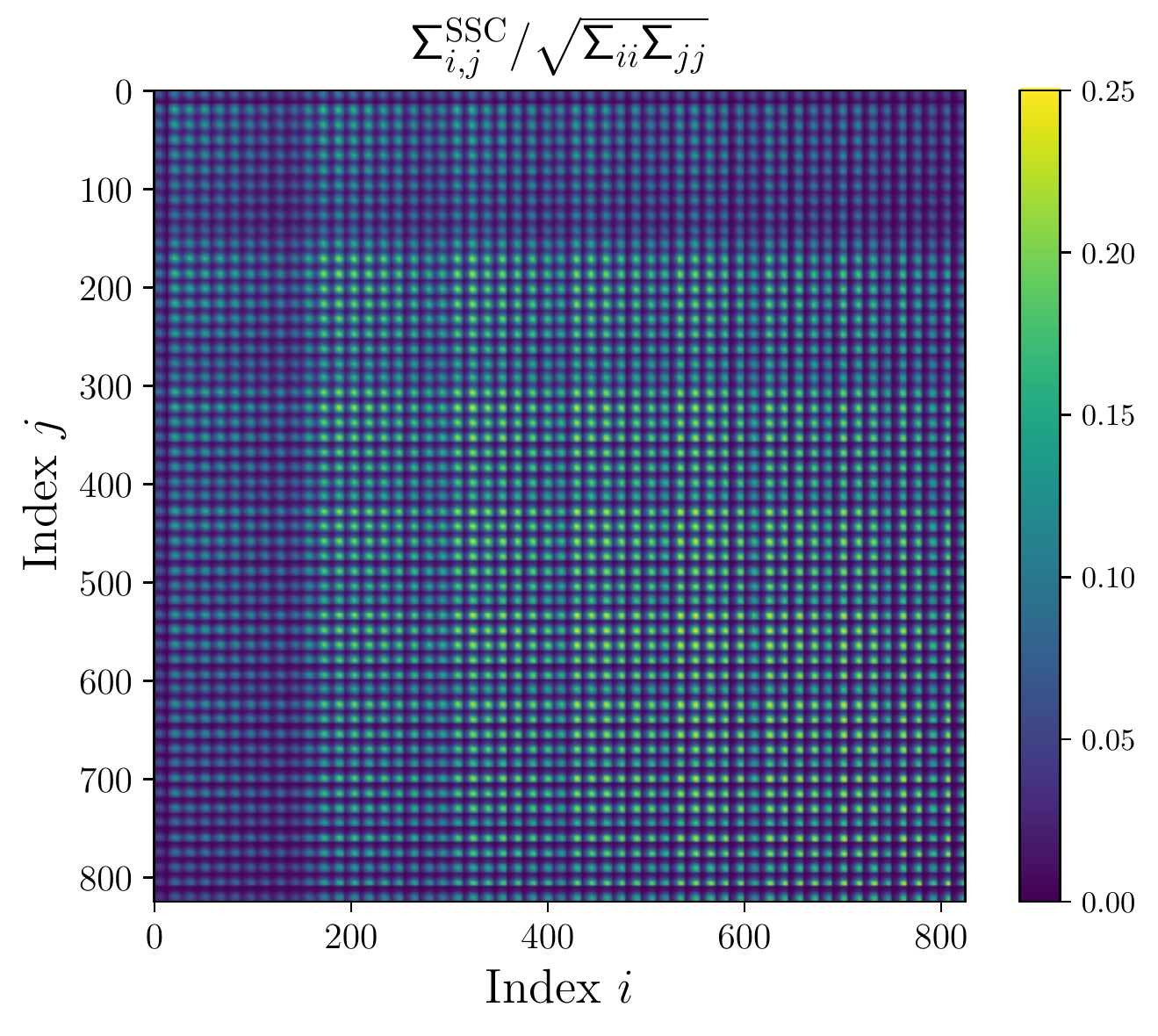}
        \caption{Contribution from the SSC term to the lensing-only part of the correlation matrix of the data vector described in Section \ref{ssec:forecasts.surveys} for a survey covering 10\% of the celestial sphere.}
        \label{fig:full_cov_lensing}
      \end{center}
    \end{figure}
    As a brief review and to introduce the notation, we provide the details of the Halo model below \cite{Li:2014sga}. In the Halo model the matter power spectrum can approximated as 
    \begin{eqnarray}
      P(k)  & = & P_{1h}(k) + P_{2h}(k)	\nonumber	\\
      & \equiv & \mathcal{I}^0_2(k) + \left[\mathcal{I}^1_1(k)\right]^2P_L(k)
    \end{eqnarray}
    where $P_{1h}(k)$ represents the contribution from correlations within a single halo and $P_{2h}(k)$ is the contribution from separate halos correlated by the linear power spectrum. Explicitly these are defined in terms of Halo model kernels $\mathcal{I}^\alpha_\beta(k)$.
    \begin{equation}
      \mathcal{I}^\alpha_\beta(k) \equiv \int dM\,\frac{dn}{dM}\,b^\alpha(M)\,\left[\frac{M}{\bar{\rho}_M}u(k|M)\right]^\beta
    \end{equation}
    where $dn/dM$ is the Halo mass function and $b(M)$ is the halo bias. Both quantities were estimated using the fits from \cite{2008ApJ...688..709T}. $u(k|M)$ is the Fourier transform of the halo density profile for a halo of mass $M$, which we model assuming a NFW profile with a concentration-to-mass relation given by \cite{Li:2014sga}.
    
    In terms of these quantities, the response of the power spectrum to the background density is given by \cite{Li:2014sga}
    \begin{align}\label{eq:response_halomod}
      &{\cal R}(k) = {\cal R}_{\rm BC}+{\cal R}_{\rm HSV}+{\cal R}_{\rm LD}\\
      &{\cal R}_{\rm BC}(k) = \frac{68}{21} \frac{P_{2h}(k)}{P(k)}\\
      &{\cal R}_{\rm HSV}(k) = -\frac{1}{3} \frac{d \log [k^3 P_{2h}(k)]}{d \log{k}} \frac{P_{2h}(k)}{P(k)}\\
      &{\cal R}_{\rm LD}(k) = I^1_2(k)
    \end{align}
    
    A few comments on these terms are in order.
    \begin{itemize}
      \item The beat coupling (BC) term is the one that has been studied extensively in the literature. It represents the fact that a short wavelength mode will grow more in the presence of a larger background density (i.e the large scale mode outside the window). 
      \item The halo sample variance (HSV) term shows the change in the number of Halos in the presence of a large scale density mode. 
      \item A large scale density mode will change the local scale factor and thus change the physical size of a mode that would otherwise be smaller. This is known as the linear dilation (LD) effect. 
    \end{itemize}
    These different contributions at $z=0$ are shown in Figure \ref{fig:Response}, and the mild redshift evolution of the overall response is presented in Figure \ref{fig:Response_z}. 

    Using this response we can compute the SSC matrix as shown in Figure \ref{fig:full_cov_lensing}. It is worth noting that, when evaluating the parameter dependence of the covariance matrix, we vary all terms entering Eq. \ref{eq:ssc_full} except for ${\cal R}$, which we keep fixed.
\end{document}